\documentclass[a4paper,11pt]{article}
\pdfoutput=1
\usepackage{jheppub}
\usepackage[T1]{fontenc}
\usepackage{bbm}
\usepackage{mathrsfs}
\usepackage{slashed}
\usepackage{caption}
\usepackage{epstopdf}
\usepackage[normalem]{ulem}
\usepackage[bottom]{footmisc}

\newcommand{\SU}[1]{\ensuremath{\mathrm{SU}(#1)}}
\newcommand{\U}[1]{\ensuremath{\mathrm{U}(#1)}}
\newcommand{\Z}[1]{\ensuremath{\mathbbm{Z}_{#1}}} 

\title{\boldmath Beyond Minimal lepton-flavored Dark Matter}

\author[a]{Mu-Chun Chen,}
\author[b]{Jinrui Huang}
\author[a]{and Volodymyr Takhistov}

\affiliation[a]{Department of Physics and Astronomy, University of California, Irvine\\
4129 Frederick Reines Hall, Irvine, CA 92617-4575, U.S.A.}
\affiliation[b]{Theoretical Division, Los Alamos National Laboratory, \\
 Los Alamos, NM 87545, U.S.A.}

\emailAdd{muchunc@uci.edu}
\emailAdd{jinruih@lanl.gov}
\emailAdd{vtakhist@uci.edu}

\preprint{UCI-TR-2015-17, LA-UR-15-27938}

\abstract{We consider a class of flavored dark matter (DM) theories where
dark matter interacts with the Standard Model lepton fields at the renormalizable level.  
We allow for a general coupling matrix 
between the dark matter and leptons whose structure is beyond the one permitted 
by the minimal flavor violation (MFV) assumption. 
It is assumed that this is the only new source of flavor violation in addition to the Standard Model (SM) 
Yukawa interactions. The setup can be described by augmenting the  SM flavor symmetry by 
an additional $\SU3_{\chi}$, under which the dark matter $\chi$ transforms.
This framework is especially phenomenologically rich, due to possible novel flavor-changing interactions
which are not present within the more restrictive MFV framework. 
As a representative case study of this setting, 
which we call ``beyond MFV'' (BMFV), 
we consider Dirac fermion dark matter which transforms
as a singlet under the SM gauge group and a triplet under $\SU3_{\chi}$. 
The DM fermion couples to the SM lepton sector through a scalar mediator $\phi$.
Unlike the case of quark-flavored DM, we show that there is no $\Z3$
symmetry within either the MFV or BMFV settings
which automatically stabilizes the lepton-flavored DM.
We discuss constraints on 
this setup from 
flavor-changing processes, DM relic abundance as well as direct and indirect detections. 
We find that relatively large flavor-changing couplings are possible, 
while the dark matter mass is still within the phenomenologically
interesting region below the TeV scale.
Collider signatures which can be potentially searched for at the lepton and hadron colliders are discussed.
Finally, we discuss the implications for decaying dark matter, 
which can appear if an additional stabilizing symmetry is not imposed.}


\begin{document} 
\maketitle
\flushbottom


\section{Introduction}
\label{sec:intro}

With the natural interpretation of dark matter (DM) 
in terms of new particles and forces,
 its existence \cite{Bertone:2004pz} is among one of the striking  
indications of new physics beyond the Standard Model (SM).
While it is now known that DM constitutes around 85\% of the matter 
in the universe \cite{Ade:2013sjv}, we are still lacking a
comprehensive understanding of its identity and properties. 
Many theories of dark matter have been put forward, where the DM
candidate is identified with a neutralino, 
axion, sterile neutrino or some other new exotic
(see \cite{Feng:2010gw} for a review).~Since
the DM interactions as well as its mass remain unknown, it
is important to explore possible new signatures as well as 
constrains on various models. 
A very motivated scenario is for dark matter to be a
weakly interacting massive particle (WIMP), since 
it was found that weakly interacting DM with a mass of around $100$ GeV
reproduces the observed relic abundance (``WIMP-miracle''). 
Such possibility has been extensively studied 
within the context of low-scale supersymmetry (SUSY), where there are natural 
DM candidates with the desired features \cite{Jungman:1995df}. 
However, with LHC 
strongly constraining minimal low-scale SUSY 
extensions of the SM \cite{Chatrchyan:2013sza,Aad:2015iea},
the above scenario becomes somewhat less appealing
and it is imperative to investigate more general or alternative frameworks.
A popular approach for analyzing generic phenomenological features 
of classes of theories is to study ``simplified models''. The simplified models  
constitute minimal extensions of the SM in terms of their particle content 
and interactions. Such approach allows for a more direct comparison between the model  
predictions and constraints from the experimental DM 
searches~\cite{Chang:2013oia,deSimone:2014pda,Cheung:2013dua}.

An interesting class of simplified DM models is associated 
with flavored dark matter (see \cite{Kile:2013ola} for a review).
Motivated by the approximate global $\SU{3}$ flavor
symmetries present in the SM, in these theories
DM is charged under the flavor symmetry,
comes in multiple copies and couples to quarks or leptons. 
While there are also other scenarios
where sizable flavor violating effects can occur, 
such as models of sneutrino DM \cite{Ellis:1983ew,Hagelin:1984wv,Ibanez:1983kw,Falk:1994es,MarchRussell:2009aq},
neutrino DM within the context of 
extra dimensions \cite{Servant:2002aq} or even more exotic possibilities \cite{Cui:2011qe}, 
flavor effects are not their main focus. 
Many studies of the flavored dark matter have concentrated on 
quark-flavored DM \cite{Agrawal:2011ze, Kile:2011mn, DiFranzo:2013vra, deSimone:2014pda}, which
leads to interesting LHC phenomenology and which usually has enhanced
direct detection potential, since a tree level coupling of DM to the target
nuclei is possible. Lepton-flavored DM \cite{Agrawal:2011ze, Bai:2014osa},
on the other hand, 
leads to very different phenomenology, whose consequences have
been less explored.

Since new contributions to flavor violating processes are very constrained,
flavored DM has been typically considered within the context of
minimal flavor violation (MFV) \cite{D'Ambrosio:2002ex}.
Within the MFV framework the only source of flavor violation are the SM Yukawa couplings, 
hence, flavor changing neutral currents (FCNC) are automatically suppressed. 
While this scenario is safe and by construction is in agreement with the
experimental constraints, it is overly restrictive for exploring new phenomenology. 
Since all forms of matter in the SM have flavor, it is worthwhile to explore
the possibility that dark matter also has its own notion of flavor. Starting with
an unrestricted flavor structure in the DM coupling to SM fermions
will allow us to quantitatively study
how much freedom in flavor transformations does the dark matter has.
In a recent work of \cite{Agrawal:2014aoa}, the SM flavor symmetries have been
extended by an additional global $\SU{3}_{\chi}$ symmetry.  
Within the model, Dirac fermion DM $\chi$ transforms as a triplet under the 
$\SU{3}_{\chi}$ and as a singlet under the SM group.
The DM $\chi$ couples to the quark fields with a new Yukawa-like coupling $\lambda$
through a scalar mediator $\phi$,
which carries electroweak charge and is a triplet under the SM color $\SU{3}_C$ group.
The main assumption of the model is that aside from the Yukawa couplings,
the only new source of flavor violation is $\lambda$. This generalizes the setup
beyond that of MFV (BMFV)
\footnote{The authors of \cite{Agrawal:2014aoa} call this Dark MFV (DMFV).}.
The coupling $\lambda$ is a priori not constrained and allows for large flavor violating
effects. On the other hand, lepton-flavored DM within BMFV has not been studied.

In this work, we consider lepton-flavored dark matter within the BMFV framework.
Specifically, the SM flavor symmetries are extended by an additional global $\SU{3}_{\chi}$, 
under which the DM transforms.
As we will show, while the lepton flavor violating processes are strongly constrained,
there exists viable parameter space
for DM to be within the several hundred GeV range.
We will also demonstrate that, contrary to
the case of quark-flavored DM within the MFV
or the BMFV, there is no $\mathbb{Z}_3$ symmetry 
available within the model to stabilize
the lepton-flavored DM. Additionally, we will discuss how 
direct and indirect detection limits the parameter space
and how DM in our scenario can naturally account for the 511 keV $\gamma$-ray excess.
Then, we comment on several possible new signatures 
at the lepton and hadron colliders. As we shall see, the lepton $e^+e^-$ colliders
provide one of the more stringent constraints for both MFV and the BMFV 
scenarios. Finally, we discuss
the implications for decaying dark matter,
which can occur if an additional
stabilizing symmetry is not imposed.

This manuscript is organized as follows. Section~\ref{sec:leptonflavor}
reviews the framework of minimal flavor violation. We then 
extend it to BMFV and present a minimal model which satisfies this hypothesis.
We comment on the assumptions of the model, what will constitute the MFV
limit and also discuss how dark matter is stabilized within this setup.
In Section~\ref{sec:constraints}, we study the constraints on the model
from lepton flavor violating processes and discuss which
structure of the new Yukawa coupling
 $\lambda$ will satisfy them, if the DM mass is to be kept
in the phenomenologically interesting region of few hundred GeV. We then discuss
constraints from direct and indirect detection 
and comment on lepton and hadron
 collider signatures in MFV and BMFV models.
The possibility of decaying dark matter is also considered.
 In Section~\ref{sec:summary},
we present combined plots that include all 
experimental constraints and summarize the work.


\section{lepton-flavored dark matter}
\label{sec:leptonflavor}

\subsection{Minimal Flavor Violation}
\label{sec:mfvreview}

We start by reviewing the framework of minimal flavor violation (MFV) \cite{D'Ambrosio:2002ex}. 
In the SM, there exists an approximate global $G_F = \U{3}^5$ symmetry \cite{Chivukula:1987py},
with the group factors $\U{3}_{Q},\U{3}_{U},\U{3}_{D}$
acting on quark fields $Q$, $U^c$, $D^c$ and the factors $\U{3}_L$, $\U{3}_{E}$
acting on the lepton fields $L, E^c$. 
The symmetry breaking is induced by the SM Yukawa couplings, which provide 
different masses to different generations of quarks and leptons.
The $G_F$ symmetry can be decomposed\footnote{Recall, that a unitary group $\U{N}$ can be separated into 
non-Abelian and Abelian components with a group product $\U{N} \simeq \SU{N} \times \U1$.}
into 
\begin{equation}
G_F \equiv \SU{3}_q^3 \times \SU{3}_{l}^2 \times \U1_B \times \U1_L \times \U1_Y \times \U1_{PQ} \times \U1_{E_R} ,
\end{equation}
where
\begin{align}
\SU{3}_q^3 &= \SU{3}_Q \times \SU{3}_U \times \SU{3}_D ,\\
\SU{3}_{l}^2 &= \SU{3}_L \times \SU{3}_E .
\end{align}
Here, the individual $\U1$ factors have been combined into the more familiar linear combinations 
of $\U1_B$, $\U1_L$, $\U1_Y$, $\U1_{PQ}$ and $\U1_{E_R}$ which
are identified with baryon $(B)$ number, lepton $(L)$ number, global hypercharge, 
Peccei-Quinn and the right handed rotation 
symmetries\footnote{The global $U(1)_{Y}$ factor coincides with the gauged $U(1)_{Y}$ in the SM. 
Hence, the ``true'' $\U1$ factor in the SM global symmetry is $\U1^4$.}.
Neglecting the $\U1$ factors,
we are left with a global $G_f = \SU{3}^5$ ``flavor symmetry''.
In the absence of the Yukawa interactions, the flavor symmetry is exact. 
The breaking of the flavor symmetry is solely parametrized by the Yukawa couplings, 
which is the only source of flavor violation within the framework of MFV.

\subsection{Beyond the MFV}
\label{sec:bmfvmodel}

Going beyond the MFV framework, the BMFV setup adds an additional ``flavor''
$\SU{3}_{\chi}$. In our analysis we shall focus on the lepton-flavored dark matter. 
The dark matter $\chi$, which we assume to be a Dirac fermion,
transforms as a triplet under the $\SU{3}_{\chi}$ symmetry
and is a singlet under the SM gauge group. 
The $\chi$ field interacts with the SM fermions 
through a mediator $\phi$, with a new Yukawa-like coupling $\lambda$.
The mediator is charged under the electroweak symmetry and transforms as a color singlet.
The coupling $\lambda$ is a priori unconstrained and large
flavor violating couplings are possible. 

In principle, the DM $\chi$ 
can interact with either the left-handed SM $\SU2_L$ doublet $l$ field
or with the right-handed SM singlet $e_R$.
For simplicity, we will concentrate on the latter case. 
Below, we present the full minimal model.
 The field content of the model is provided in Table~\ref{fig:DMmodel},
where we have also included the quark sector for completeness.
 Since a Majorana DM mass term would violate the $\SU3_{\chi}$ 
symmetry, it is forbidden.
\begin{figure}[htb]
\centering
\begin{tabular}{|c||ccc|cccccc|} 
\hline
 & $\SU3_c$ & $\SU2_L$ & $\U1_Y$ & $\SU3_Q$ & $\SU3_U$ & $\SU3_D$ & $\SU3_L$ & $\SU3_E$  & $\SU3_{\chi}$\\
\hline
\hline
$Q_L$ & $\mathbf{3}$ & $\mathbf{2}$ & 1/6 & $\mathbf{3}$ & $\mathbf{1}$ & $\mathbf{1}$ & $\mathbf{1}$ & $\mathbf{1}$ & $\mathbf{1}$ \\ 
$u_R$ & $\mathbf{3}$ & $\mathbf{1}$ & 2/3 & $\mathbf{1}$ & $\mathbf{3}$ & $\mathbf{1}$ & $\mathbf{1}$ & $\mathbf{1}$ & $\mathbf{1}$\\ 
$d_R$ & $\mathbf{3}$ & $\mathbf{1}$ & -1/3 & $\mathbf{1}$ & $\mathbf{1}$ & $\mathbf{3}$ & $\mathbf{1}$ & $\mathbf{1}$ & $\mathbf{1}$\\ 
\hline
$L_L$ & $\mathbf{1}$ & $\mathbf{2}$ & -1/2 & $\mathbf{1}$ & $\mathbf{1}$ & $\mathbf{1}$ & $\mathbf{3}$ & $\mathbf{1}$ & $\mathbf{1}$ \\ 
$e_R$ & $\mathbf{1}$ & $\mathbf{1}$ & -1 & $\mathbf{1}$ & $\mathbf{1}$ & $\mathbf{1}$ & $\mathbf{1}$ & $\mathbf{3}$ & $\mathbf{1}$  \\ 
\hline
$H$ & $\mathbf{1}$ & $\mathbf{2}$ & 1/2 & $\mathbf{1}$ & $\mathbf{1}$ & $\mathbf{1}$ & $\mathbf{1}$ & $\mathbf{1}$ & $\mathbf{1}$ \\ 
\hline
$\phi$ & $\mathbf{1}$ & $\mathbf{1}$ & -1 & $\mathbf{1}$ & $\mathbf{1}$ & $\mathbf{1}$ & $\mathbf{1}$ & $\mathbf{1}$ & $\mathbf{1}$  \\ 
$\chi$ & $\mathbf{1}$ & $\mathbf{1}$ & 0 & $\mathbf{1}$ & $\mathbf{1}$ & $\mathbf{1}$ & $\mathbf{1}$ & $\mathbf{1}$ & $\mathbf{3}$ \\ 
\hline
\hline
$Y_{U}$ & $\mathbf{1}$ & $\mathbf{1}$ & 0 & $\mathbf{3}$ & $\mathbf{\overline{3}}$ & $\mathbf{1}$ & $\mathbf{1}$ & $\mathbf{1}$ & $\mathbf{1}$  \\ 
$Y_{D}$ & $\mathbf{1}$ & $\mathbf{1}$ & 0 & $\mathbf{3}$ & $\mathbf{1}$ & $\mathbf{\overline{3}}$ & $\mathbf{1}$ & $\mathbf{1}$ & $\mathbf{1}$  \\ 
$Y_{E}$ & $\mathbf{1}$ & $\mathbf{1}$ & 0 & $\mathbf{1}$ & $\mathbf{1}$ & $\mathbf{1}$ & $\mathbf{3}$ & $\mathbf{\overline{3}}$ & $\mathbf{1}$  \\ 
$\lambda$ & $\mathbf{1}$ & $\mathbf{1}$ & 0 & $\mathbf{1}$ & $\mathbf{1}$ & $\mathbf{1}$ & $\mathbf{1}$ & $\mathbf{3}$ & $\mathbf{\overline{3}}$ \\
\hline 
\end{tabular}
\captionof{table}{Field content of the minimal lepton-flavored BMFV model.}
\label{fig:DMmodel}
\end{figure}

The complete renormalizable Lagrangian of our model is given by
\begin{align} 
\mathscr{L} = \mathscr{L}_{SM} &+ i \overline{\chi} \slashed{\partial} \chi - m_{\chi} \overline{\chi} \chi  
- (\lambda_{i j} \overline{e}_{R_i} \chi_{j} \phi + \text{h.c.})  \\
&+ (D_{\mu} \phi)^{\dagger} (D^{\mu} \phi) - m_{\phi}^2 \phi^{\dagger} \phi 
+ \lambda_{H \phi} (\phi^{\dagger} \phi)( H^{\dagger} H) + \lambda_{\phi \phi} (\phi^{\dagger} \phi)^2 , \nonumber
\end{align}
where in order to systematically account for the flavor violating effects, 
the SM charged lepton Yukawa coupling $Y_{E}$ as well as $\lambda$
have been treated as spurion fields, 
which also transform under the symmetry group (see Table~\ref{fig:DMmodel}).

\subsection{Model assumptions, structure of $\lambda$ and the MFV limit}
\label{sec:assumptions}

We will now specify  what constitutes the ``MFV limit'' within the BMFV framework.
For the case of lepton-flavored DM, to recover the MFV from BMFV,
one simply has to identify the $\SU{3}_{\chi}$ with
either of the SM leptonic flavor symmetries, $\SU{3}_E$ or the $\SU{3}_L$. 
The parameters of the model will be restricted, in order to enforce
consistency with the MFV assumptions \cite{Agrawal:2011ze}.

First, consider the case where $\SU{3}_{\chi}$ is identified with
$\SU{3}_E$. The MFV expansion of the DM coupling matrix $\lambda$ and the 
$\chi$ mass matrix $m$, is given in terms of the SM charged lepton Yukawa couplings $Y_{E}$ as,  
\begin{align}
\lambda_{ij}^{\text{MFV}} & =(\alpha \mathbf{1} + \beta~ [Y^{\dagger}_{E} Y_{E}])_{ij} , \\
m_{{ij}}^{\text{MFV}} & = (m_{\chi} \mathbf{1} + \Delta_m~ [Y^{\dagger}_{E} Y_{E}])_{ij} ,
\end{align}
where we have kept only the lowest order term in the expansion 
and $\alpha$, $\beta$, $m_{\chi}$ and $\Delta_m$ 
are the expansion parameters. 
On the other hand, if $\SU{3}_{\chi}$ is identified with $\SU{3}_L$,
then $\lambda_{i j}^{\text{MFV}}$ will simply reduce to the
charged lepton Yukawa coupling $\lambda_{i j}^{\text{MFV}} = Y_{e}$, while the
mass term expansion will stay the same as above.
It is assumed that any such expansions are convergent.

To make the discussion concrete, one may choose the basis where the observed flavor mixing 
in the SM is attributed to the flavor rotation in one sector only. For example, in the quark sector, 
a common choice for the Yukawa matrices constitutes  
$Y_D = \lambda_D$ and $Y_U = V_{CKM} \lambda_U$, which automatically satisfies the more stringent 
flavor constraints in the down quark sector. 
Here, $\lambda_D = (y_d, y_s, y_b)$ and $\lambda_U = (y_u, y_c, y_t)$
are diagonal and $V_{CKM}$ is the CKM matrix.
In this parametrization, flavor violating contributions will only appear through
the off-diagonal CKM components, which will be present in the 
MFV expansion of the model parameters due to the $[Y_U^{\dagger} Y_U]$ terms.
Similarly, in the lepton sector, it is commonly chosen that the charged lepton Yukawa 
coupling $Y_{E}$ is diagonal, with $Y_{E} = \lambda_E = (y_e, y_{\mu}, y_{\tau})$.
Hence, $Y_{E}$ will not provide any flavor violating contributions in the MFV expansion.
Thus far, due to large uncertainty, we have omitted mentioning neutrino masses.
Assuming that neutrinos are Dirac fermions, 
the above discussion of quark Yukawas will directly translate to the lepton sector.
With a diagonal $Y_{E}$, all the lepton flavor violating contributions
will come from the off-diagonal elements of the PMNS mixing matrix, which will appear in the 
MFV expansion through the $[Y_{\nu}^{\dagger} Y_{\nu}]$ terms.
The situation is slightly different if neutrinos are Majorana, generally discussed
within the context of the see-saw mechanism \cite{Cirigliano:2005ck}.
For the remainder of this work, we will neglect the possible 
flavor violating contributions from the lepton sector, since their
size is negligible compared to the possible contributions
from the new DM coupling $\lambda$.

In the case of BMFV, unlike the MFV, coupling $\lambda$
is a priori not restricted in its form. Hence, both the diagonal (flavor-preserving)
and the off-diagonal (flavor-changing)
coefficients, can potentially be of $\mathcal{O}(1)$.
Neglecting the SM Yukawa contributions,
the mass term expansion is now given 
in terms of $\lambda_{i j}$ as
\begin{equation}
m_{i j}^{\text{BMFV}} = (m_{\chi} \mathbf{1} + \Delta_m~ [\lambda^{\dagger} \lambda] )_{i j} .
\end{equation}
Further, since there will be induced mass-splitting among the components in the DM multiplet 
induced by the RGEs, $\Delta_m$ is estimated to be~\cite{Agrawal:2014aoa}
\begin{equation} \label{eq:mass_split}
\Delta_m  \sim \dfrac{1}{16 \pi^2} \log \Big( \dfrac{m_{\chi}^2}{\Lambda^2} \Big) ,
 \end{equation}
where $\Lambda$ is the UV-scale at which $\SU{3}_{\chi}$ gets broken.
Detailed discussion regarding possible UV-completions
 of such scenario is beyond the scope of this work.
Throughout our analysis, unless otherwise stated,
we will assume that the three components in the DM triplet 
$\chi$ are highly degenerate in mass with $m_{a} \simeq m_{\chi}$ for $a =1, 2, 3$.

\subsection{Dark matter stability}
\label{sec:dmstability}

To ensure DM stability, it is common to impose
an additional symmetry which forbids DM from
decaying. In \cite{Batell:2011tc}, it was pointed out that 
within the MFV framework, assuming quark-flavored DM,
there is a residual $\Z3$ symmetry which automatically stabilizes 
the DM\footnote{Recall that $\Z{N}$ is the center of the $\SU{N}$ group. The residual 
$\Z3$ symmetry in the quark-flavored DM case is thus related to the centers of the $SU(3)$'s present in the model.}.
The argument was later extended~\cite{Agrawal:2014aoa} to the BMFV framework assuming quark-flavored DM. 
It is then natural to wonder, if there is an analogous argument,
in the context of either MFV or the BMFV framework,
for the case of lepton-flavored DM. As we will show below,
there is no residual discrete symmetry to automatically stabilize the DM for either of the 
two of lepton-flavored DM scenarios.

Let us first discuss the more general BMFV framework.
For the lepton-flavored DM, the most general invariant operator involving the DM triplet $\chi$, mediator $\phi$ 
as well as the SM fields and the Yukawa spurions, 
which parametrize flavor symmetry breaking, is
\begin{align}
\mathcal{O}_{DM} \sim~& \chi \dots \overline{\chi} \dots \phi \dots \phi^{\dagger} \dots
L \dots \overline{L} \dots e_{R} \dots \overline{e}_{R} \dots  Y_E \dots Y_E^{\dagger} \dots \lambda \dots \lambda^{\dagger} 
\dots G  , 
\end{align}
where the dots signify an arbitrary number of insertions of a given field type and $G$  
schematically denotes the field combination
which is itself invariant under the color $\SU{3}_c$ and which renders 
$\mathcal{O}_{DM}$ invariant under the electroweak symmetry.
Since the fields transform in fundamental
and anti-fundamental representations of the flavor
 $\SU{3}$ symmetries, the product\footnote{Recall, the $\SU{3}$ singlets can be formed from 
the fundamental $\mathbf{3}$ and anti-fundamental $\overline{\mathbf{3}}$ representations via 
multiples of $\mathbf{3} \times \mathbf{3} \times \mathbf{3}$, 
$\overline{\mathbf{3}} \times \overline{\mathbf{3}} \times \overline{\mathbf{3}}$
or $\mathbf{3} \times \overline{\mathbf{3}}$.} 
$\mathcal{O}_{DM}$ results in an $\SU{3}$ singlet only if the number of 
the corresponding $\SU{3}$ triplets $\mathbf{3}$ minus the number
of the corresponding anti-triplets $\overline{\mathbf{3}}$ is 0 modulo 3.
This results in a set of conditions, one for each $\SU{3}$ symmetry, which must
hold for $\mathcal{O}_{DM}$ to be an invariant
\begin{align}
  (N_L - N_{\overline{L}} + N_{Y_{E}} - N_{Y_{E}^{\dagger}})
 \mod 3  &= 0,~~~\text{for}~ \SU{3}_L , \\
 (N_E - N_{\overline{E}} - N_{Y_{E}} + N_{Y_{E}^{\dagger}} 
+ N_{\lambda} - N_{\lambda^{\dagger}}) \mod 3  &= 0,~~~\text{for}~ \SU{3}_E , \\
 (N_{\chi} - N_{\overline{\chi}} 
- N_{\lambda} + N_{\lambda^{\dagger}}) \mod 3  &= 0,~~~\text{for}~ \SU{3}_{\chi} ,
\end{align}
where $N_j$ denotes the number of insertions of the field $j$. 
Canceling the dependency on $Y_{E}$ and $\lambda$,
we obtain
\begin{equation} \label{eq:residualZ3}
(N_{\chi} - N_{\overline{\chi}}
 + N_{L} - N_{\overline{L}} 
 + N_{E} - N_{\overline{E}}) \mod 3  = 0 ,
\end{equation}
which can be interpreted as a condition on the field charges
transforming under a $\Z{3}$ symmetry.

Consider now the case with
$N_{\chi} = 1$, $N_{\overline{\chi}} = N_{\phi} = N_{\phi^{\dagger}} = 0$, 
which corresponds to operators that lead to dark matter decay. Since the $\Z{3}$ invariance condition
\eqref{eq:residualZ3} can still be satisfied, provided an appropriate combination
of the lepton fields, the dark matter stability is not ensured.
This argument applies identically in the MFV framework, 
where $\SU{3}_{\chi}$ is identified with either $\SU{3}_E$
or $\SU{3}_L$. Since in the quark-flavored 
case of \cite{Batell:2011tc,Agrawal:2014aoa} there was an extra 
condition coming from the additional color $\SU{3}_c$, the final equation for $\Z{3}$
was further reduced. In that case, the above condition
only has dependency on the DM $\chi$ and mediator $\phi$ fields and not on the SM fermions,
enabling the $\Z{3}$ to stabilize the dark matter. 

Thus, we have shown that the $\Z{3}$ symmetry which stabilizes
quark-flavored DM, in the MFV and BMFV settings, is not available
for the lepton-flavored DM case. Hence, we will assume
that there is an additional $\Z{2}$ which stabilizes the DM. 
Under this $\Z{2}$ symmetry the $\chi$ and $\phi$ fields are odd while the SM 
fields are even and the dark matter decay operators are consequently forbidden.
The case of decaying dark matter will be discussed in Section~\ref{ssec:dmdecay}.


\section{Constraints and signals}
\label{sec:constraints}

Below, we discuss constraints as well as possible signals from the various
sources related to lepton BMFV as specified in Sec. \ref{sec:bmfvmodel}. 

\subsection{Flavor constraints}
\label{sec:flavorconstraints}

\subsubsection{Lepton flavor violating processes}
\label{sec:lfvprocess}

\begin{figure}[htb] 
    \centering
	\normalsize
\includegraphics[width=0.4\linewidth]{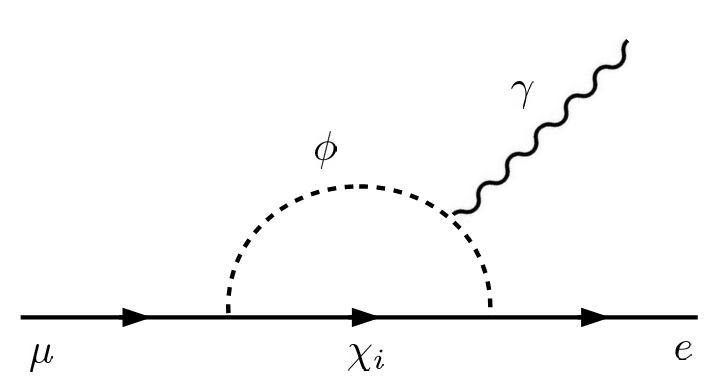}
\caption{Contributing diagram for $\mu \rightarrow e\gamma$.}
\label{fig:mu2egamma}
\end{figure}

As the structure of $\lambda$ in the BMFV  is a priori unrestricted,
unlike the case of MFV, unsuppressed lepton flavor-violating (LFV)
processes are possible and will occur through the off-diagonal elements of $\lambda$. 
Assuming the flavor violating couplings $\lambda_{i j~ (i \neq j)}$ are sizable,  there are strong constraints on 
mediator mass $m_{\phi}$ and dark matter mass $m_{\chi}$ 
from the LFV processes. As we will see, these LFV processes give rise to the most severe constraints on our model in some parameter space. 
Lepton flavor violating processes have been extensively studied,
as shown by the stringency of the current experimental limits as well as the
expected sensitivities 
of future experiments, which can be found in~\cite{Adam:2013mnn, Baldini:2013ke, Aubert:2009ag, Aushev:2010bq, Bellgardt:1987du, Blondel:2013ia, Hayasaka:2010np, Dohmen:1993mp, Kuno:2005mm, Bertl:2006up, Natori:2014yba}. 
Within BMFV, processes such as $\mu \rightarrow e\gamma$
will arise at the 1-loop level, as displayed in Figure~\ref{fig:mu2egamma}.
The effective amplitude for this process can be expressed as 
\begin{equation} \label{eq:mu2egamma_full}
\mathcal{M}_{\mu \rightarrow e\gamma} = 
\dfrac{e}{2 m_{\mu}} \epsilon^{\ast \alpha} \overline{u}_e
[i \sigma_{\beta \alpha} q^{\beta} (a_R^{(\mu e \gamma)} P_L + a_L^{(\mu e \gamma)} P_R) ] u_{\mu}  ,
\end{equation}
where $P_{R,L} = (1 \pm \gamma_5)/2$ are the projection operators, 
$\sigma^{\mu\nu} \equiv \dfrac{i}{2} [\gamma^{\mu}, \gamma^{\nu}]$, 
the $\overline{u}_e, u_\mu$ are the spinors that satisfy the Dirac equation,
$e$ is the electric charge, $\epsilon$ is the photon polarization vector
and the indices $L, R$ in 
$a_L^{(\mu e \gamma)}, a_R^{(\mu e \gamma)}$ refer to the  
electron chirality\footnote{We use the notation of \cite{Kersten:2014xaa}, 
which has indices
interchanged compared to the usual notation
(see for example textbook of Cheng and Li).}.
This leads to the branching ratio of
\begin{equation}
\mathcal{B} (\mu \rightarrow e\gamma) = \dfrac{3 \pi^2 e^2}{ G_F^2 m_{\mu}^4} 
(|a_L^{ \mu e \gamma)}|^2 + |a_R^{(\mu e \gamma)}|^2) .
\end{equation}
With the current 90\% C. L. experimental limit on the branching ratio being
\begin{equation}
\mathcal{B} (\mu \rightarrow e\gamma) \equiv 
\Gamma(\mu \rightarrow e\gamma) / \Gamma(\mu_{\text{total}}) < 5.7 \times 10^{-13} ,
\end{equation}
$\mu \rightarrow e\gamma$ is 
one of the most 
constrained processes 
among 
various LFV channels.
We will thus focus our attention on this decay mode.

Within the SM \cite{Petcov:1976ff, Ma:1980gm},
various LFV processes are extremely suppressed, with the predicted
rate for $\mu \rightarrow e\gamma$ being of the order of $10^{-56}$
and arising through neutrino mixing. On the other hand,
going beyond the SM, sizable 
contributions to these channels often are present. 
This is also the case for the MSSM,
within the context of which $\mu \rightarrow e\gamma$ 
has been extensively studied \cite{Hisano:1995cp}.
As we discuss, the analytic result for $\mu \rightarrow e\gamma$ for the BMFV
setting can be obtained directly from the MSSM,
if the couplings are appropriately 
identified between the two frameworks. 
The contributing $\mu \rightarrow e\gamma$ diagrams in the MSSM
come from loops that involve sfermions and charginos/neutralinos. 
For the BMFV scenario, we are interested in the sfermion-neutralino loop,
which contains a neutral fermion and a charged scalar and which 
we can interpret in terms of $\chi$ and $\phi$. 
We note, that in order to obtain the
flavor violation within the MSSM,
mass insertion terms must be present in the loop
(see for example \cite{Hisano:1995nq}).
This is also the case for the MFV, 
where the couplings are nearly flavor diagonal.
On the other hand, in the BMFV, the off-diagonal $\lambda$ couplings
mediate this process directly. Taking the MSSM sfermion-neutralino
loop and neglecting the mass insertion term \cite{Kersten:2014xaa},
the amplitude is
\begin{equation} \label{eq:mu2egamma_calc}
a_R^{(\mu e \gamma)} = - \dfrac{m_{\mu}^2}{192 \pi^2 m_{\phi}^2} 
\Big[\sum_{i = 1}^3 (\lambda_{1 i}^{\ast} \lambda_{2 i})\Big]
F_1 (x) ,
\end{equation}
where $x = m_{\chi}^2 / m_{\phi}^2$ and $\lambda_{1 i}^{\ast}, \lambda_{2 i}$ correspond
to the electron and muon couplings of $\chi_i$ running in the loop.
The form factor $F_1 (x)$ is given by
\begin{equation}
F_1 (x) = \dfrac{2}{(1 - x)^4} \Big[ 1 - 6x + 3x^2 + 2x^3 - 6x^2 \log x \Big] .
\end{equation}
Labeling the components of the triplet field 
as $\chi_1, \chi_2, \chi_3$, 
the first two particles contribute to the amplitude above with one diagonal
and one off-diagonal coupling. On the other hand, the $\chi_3$ 
contribution originates from purely flavor-violating off-diagonal $\lambda$ elements.

\begin{figure}[t]
\centering
\begin{minipage}{.5\textwidth}
  \centering
  \includegraphics[width=1\linewidth]{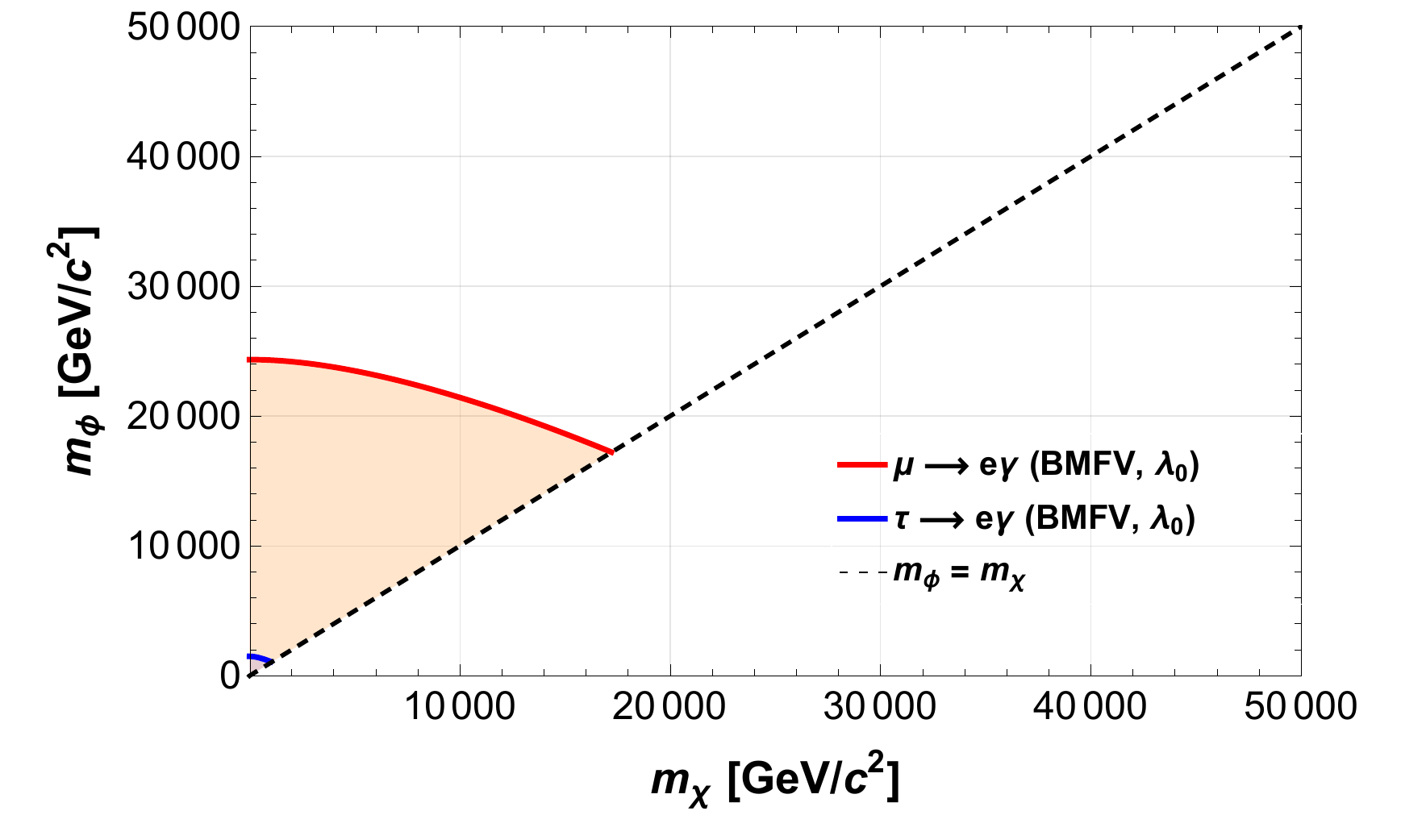}
\end{minipage}%
\begin{minipage}{.5\textwidth}
  \centering
  \includegraphics[width=1\linewidth]{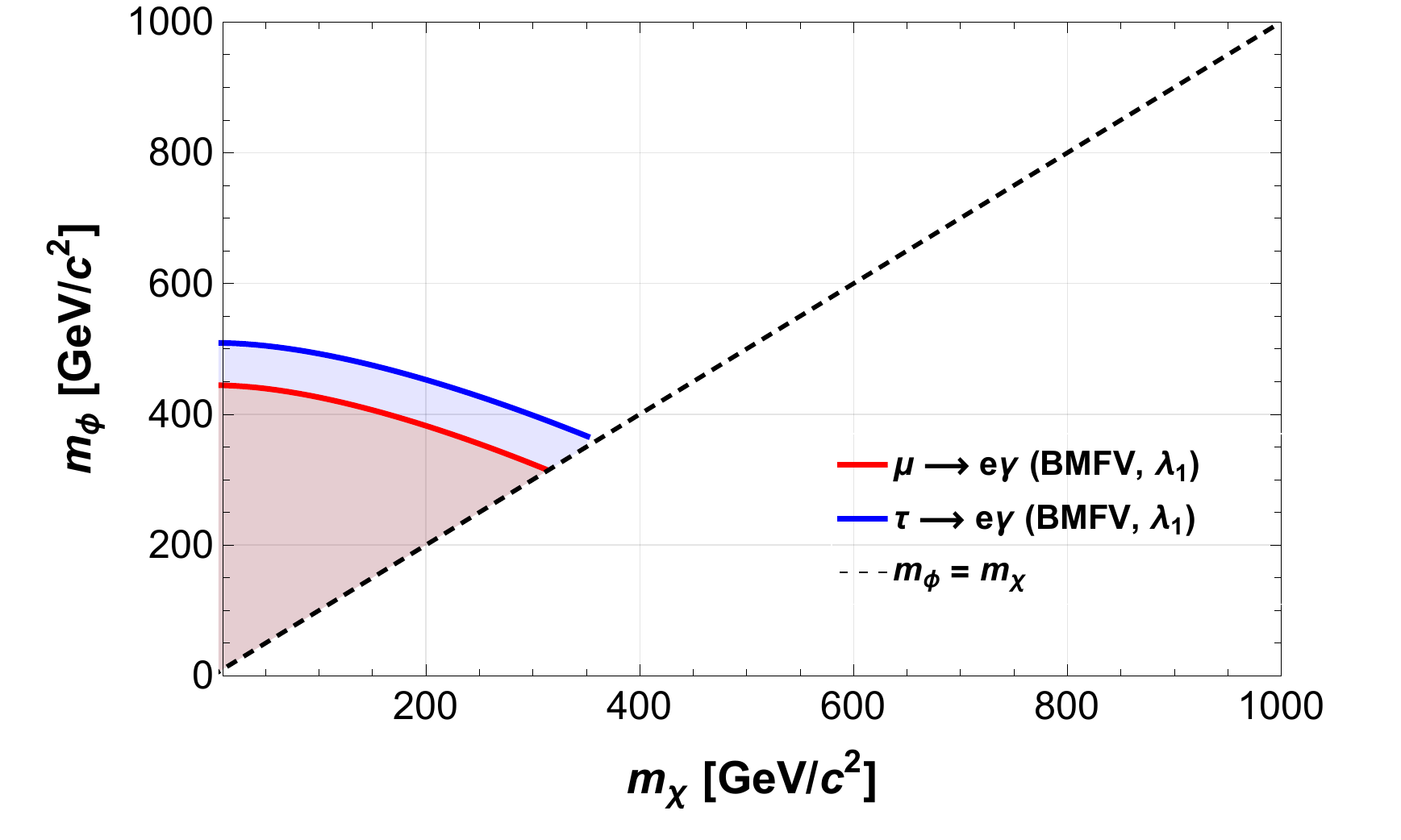}
\end{minipage}
\caption{The unshaded region above the dashed line indicates the allowed parameter space on the $m_{\phi} - m_{\chi}$ plane by the LFV constraints, 
for the choice of DM coupling matrix
$\lambda = \lambda_0$ (left)
and $\lambda = \lambda_1$ 
(right). 
}
\label{fig:mu2egamma_param}
\end{figure}

If all $\lambda$ couplings 
are taken to be 1, 
\[ \lambda_0 = 
\left( \begin{array}{ccc}
1 & 1 & 1 \\
1 & 1 & 1 \\
1 & 1 & 1 \\
\end{array} \right) \; ,
\]
the lower limits on 
the DM mass $m_{\chi}$ and the mediator mass $m_{\phi}$ 
from the experimental constraint on $\mu \rightarrow e\gamma$  
will lie in the uninteresting region of 10-50 TeV 
(see Figure~\ref{fig:mu2egamma_param}). 
In order to have DM mass within the 
phenomenologically interesting 
parameter space of several hundred GeV, some of the couplings will
have to be suppressed. One possible choice,
is to assume that the diagonal $\lambda$ couplings are 
dominant and are equal to 1, which will allow
to make a more direct comparison to the MFV scenario.
In this case, to maximize the LFV contributions,
we set $\lambda_{12} = \lambda_{21} = 0$.   
This ensures
that only $\chi_3$ mediates the process.
Taking the LFV couplings of $\chi_3$ to be 
$\lambda_{2 3} = \lambda_{1 3} = 10^{- \frac{3}{2}}$,
places the DM mass below the TeV scale as desired.

Thus far, we have discussed six of the nine $\lambda$ couplings,
which are involved 
in $\mu \rightarrow e\gamma$.
We can similarly restrict the remaining three couplings, 
which describe the $\tau$ lepton, by looking
at the constraints on $\tau \rightarrow e\gamma$
and $\tau \rightarrow \mu\gamma$. 
Requiring that the DM mass is 
in the range of a few hundred GeV,
the remaining $\tau$ couplings are set to be $10^{- \frac{1}{2}}$ 
while satisfying experimental constraints on all LFV processes. 
The full structure of matrix $\lambda$ for these constraints
is then given by
\[ \lambda_1 = 
\left( \begin{array}{ccc}
1 & 0 & 10^{- \frac{3}{2}} \\
0 & 1 & 10^{- \frac{3}{2}} \\
10^{- \frac{1}{2}} & 10^{- \frac{1}{2}} & 1 
\end{array} \right) .
\] 
In Figure~\ref{fig:mu2egamma_param},
we show the constraints from LFV
processes for two representative choices for the structure of $\lambda$,
the case $\lambda = \lambda_0$, where all $\lambda$ couplings are taken to be 1,  
as well as the case 
of $\lambda = \lambda_1$.
We stress, that the above choice of $\lambda_1$ is not unique. 
Suppressing the diagonal elements would 
allow to enlarge the off-diagonal LFV couplings. 
As an example, consider the coupling matrix $\lambda_2$, whose
structure is given by
\[ \lambda_2 = 
\left( \begin{array}{ccc}
0 & 1 & 10^{- \frac{3}{2}} \\
1 & 0 & 10^{- \frac{3}{2}} \\
10^{- \frac{1}{2}} & 10^{- \frac{1}{2}} & 0 
\end{array} \right) .
\]
The matrix $\lambda_2$ is a variation of $\lambda_1$, 
but with the diagonal entries set to 0 and off-diagonal $\lambda_{12}$ and $\lambda_{21}$
terms set to 1. The structure of $\lambda_2$ is manifestly  
non-MFV. Since constraints on $\lambda_2$ are similar to those on $\lambda_1$, 
we will focus only on $\lambda_1$ for the remainder of this work.

\subsubsection{Muon $(g - 2)$}
\label{sec:muong2}

The measured value \cite{Hanneke:2008tm} of the magnetic moment of electron
is in a good agreement with the SM predictions \cite{Gabrielse:2006gg}. 
On the other hand, 
the measured value of the muon's magnetic moment\cite{Bennett:2006fi, Roberts:2010cj}
\begin{equation}
a_{\mu}^{\text{exp}} \equiv \dfrac{g_{\mu} - 2}{2} \equiv \dfrac{\mu_{\mu}}{(e \hbar / 2 m_{\mu})} - 1 
= (11659208.9 \pm 6.3) \times 10^{-10} ,
\end{equation}
differs from the SM calculations \cite{Hagiwara:2011af, Blum:2013xva} 
of $a_{\mu}^{\text{SM}} = (11659182.8 \pm 4.9) \times 10^{-10}$
by 
\begin{equation}
a_{\mu}^{\text{diff}} = a_{\mu}^{\text{exp}} - a_{\mu}^{\text{SM}} = (26.1 \pm 8.0) \times 10^{-10} ,
\end{equation} 
which corresponds to  
around $3 \sigma$ deviation. 
Proposed future experiments \cite{Grange:2015eea} 
aim at improving the precision to $1.6 \times 10^{-10}$ (0.14 ppm).

While one cannot make a definite  
statements about this 
discrepancy at present, it can still be of interest to explore it further by 
interpreting it as a possible sign of new physics. 
This issue has been extensively studied in the MSSM
\cite{Moroi:1995yh,Carena:1996qa,Martin:2001st,Chacko:2001xd}, 
with the sfermion or gaugino running in the loop. 
Its effective amplitude is obtained by identifying 
$a_{L}^{(\mu e\gamma)} = a_{R}^{(\mu e\gamma)} = a_{\mu}$ in the effective amplitude for  
$\mu \rightarrow e\gamma$ in Equation~\eqref{eq:mu2egamma_calc}, 
\begin{equation} \label{eq:mumom_full}
\mathcal{M}_{\mu} = 
\dfrac{e}{2 m_{\mu}} \epsilon^{\alpha} \overline{u}_{\mu}
[i \sigma_{\beta \alpha} q^{\beta} a_{\mu}] u_{\mu}  .
\end{equation} 
It is worthy to note, that while the loop form factor will stay the same
for both the magnetic moment and $\mu \rightarrow e\gamma$, 
in the MSSM the 
two amplitudes are distinct. 
In MSSM, a 
mass insertion terms is required to be present for the flavor-changing process, $\mu \rightarrow e\gamma$, 
but not for the magnetic moment as it is flavor-conserving. This statement also holds for 
the case  
of flavored DM models with the MFV assumption. 
On the other hand, within BMFV, the results for the
two processes will have the same general analytic structure, 
differing only in the couplings. 
Hence, by properly replacing the coupling constants in the MSSM loop contribution due to 
slepton and neutralino~\cite{Martin:2001st},
the muon magnetic moment 
in our model with BMFV can be obtained,
\begin{equation} \label{eq:anom_mom}
\delta a_{\mu} = - \dfrac{m_{\mu}^2}{192 \pi^2 m_{\phi}^2} 
\Big[\sum_{i = 1}^3 (\lambda_{2 i}^{\ast} \lambda_{2 i})\Big]
F_1 (x) ,
\end{equation}
where $x$ and $F_1 (x)$ are the same as in Equation~\eqref{eq:mu2egamma_calc}.
Since in the BMFV case with a general coupling matrix $\lambda = \lambda_{0}$, 
all three components $\chi_{i}$ can couple to the muon,    
there will be multiplied  
contributions to the muon magnetic moment
compared to the MFV case \cite{Agrawal:2014aoa}
or its ``lepto-philic'' DM \cite{Chang:2014tea} variation. 
However, all the extra contributions will come with a negative sign 
and cannot account for the $a_{\mu}^{\text{diff}}$
and the discrepancy only increases. 
Thus, we shall not discuss this process further.

\subsection{Relic abundance}
\label{sec:relic}

Assuming that $\chi$ is a thermal WIMP dark matter, the relic abundance is found
from the annihilation rate to the SM particles. 
The dominant contribution comes from the $t$-channel,
through a mediator $\phi$ exchange, resulting in two leptons.
The relic abundance calculation in the case of MFV was already 
described in \cite{Agrawal:2011ze, Bai:2014osa}.
Since in the BMFV case, each of the DM particles $\chi_{i}$ can couple
to the $e, \mu$ and $\tau$ leptons, 
additional contributions become possible.

For a general case with multiple DM species of distinct masses, 
the relic density is  
determined by the self-annihilation process, 
$\chi_a \overline{\chi}_a \rightarrow l_{i} \overline{l}_{j}$, 
where $i, j$ label the lepton generation and $a$ labels the DM type. 
Following the MFV calculation \cite{Agrawal:2011ze, Bai:2014osa}
the $s$- and $p$-wave annihilation cross-section 
is given by 
\begin{align}
\label{eq:annihilationX}
\dfrac{1}{2} \langle \sigma v \rangle_{\chi_a \overline{\chi}_a} = &~\dfrac{1}{2} 
 \Bigg[ \dfrac{(\lambda_{ia}^{\ast} \lambda_{ja} \lambda_{ia} \lambda_{j a}^{\ast}) m_{\chi_{a}}^2}{32 \pi (m_{\chi_{a}}^2 + m_{\phi}^2)^2}  \\
& +  v^2 \dfrac{(\lambda_{ia}^{\ast} \lambda_{ja} \lambda_{ia} \lambda_{ja}^{\ast}) m_{\chi_{a}}^2 (-5 m_{\chi}^4 - 18 m_{\chi_{a}}^2 m_{\phi}^2 
+ 11 m_{\phi}^4)}{768 \pi (m_{\chi_{a}}^2 + m_{\phi}^2)^4} \Bigg]  \nonumber \\
= &~ s + p v^2 , \nonumber 
\end{align}
where $v$ is relative velocity of DM particles ($\sim 0.3 c$ at freeze-out).
We review the standard procedure for calculating
the relic abundance in Appendix \ref{app:calcrelic}.
Since within the BMFV the flavor-violating off-diagonal couplings
$\lambda_{i j~(i \neq j)}$ can contribute,
the total annihilation cross-section will be enhanced.  
Due to the velocity suppression,  
we consider only the $s$-wave contribution at the freeze-out.
\begin{figure}[t]
\centering
\begin{minipage}{.5\textwidth}
  \centering
  \includegraphics[width=1\linewidth]{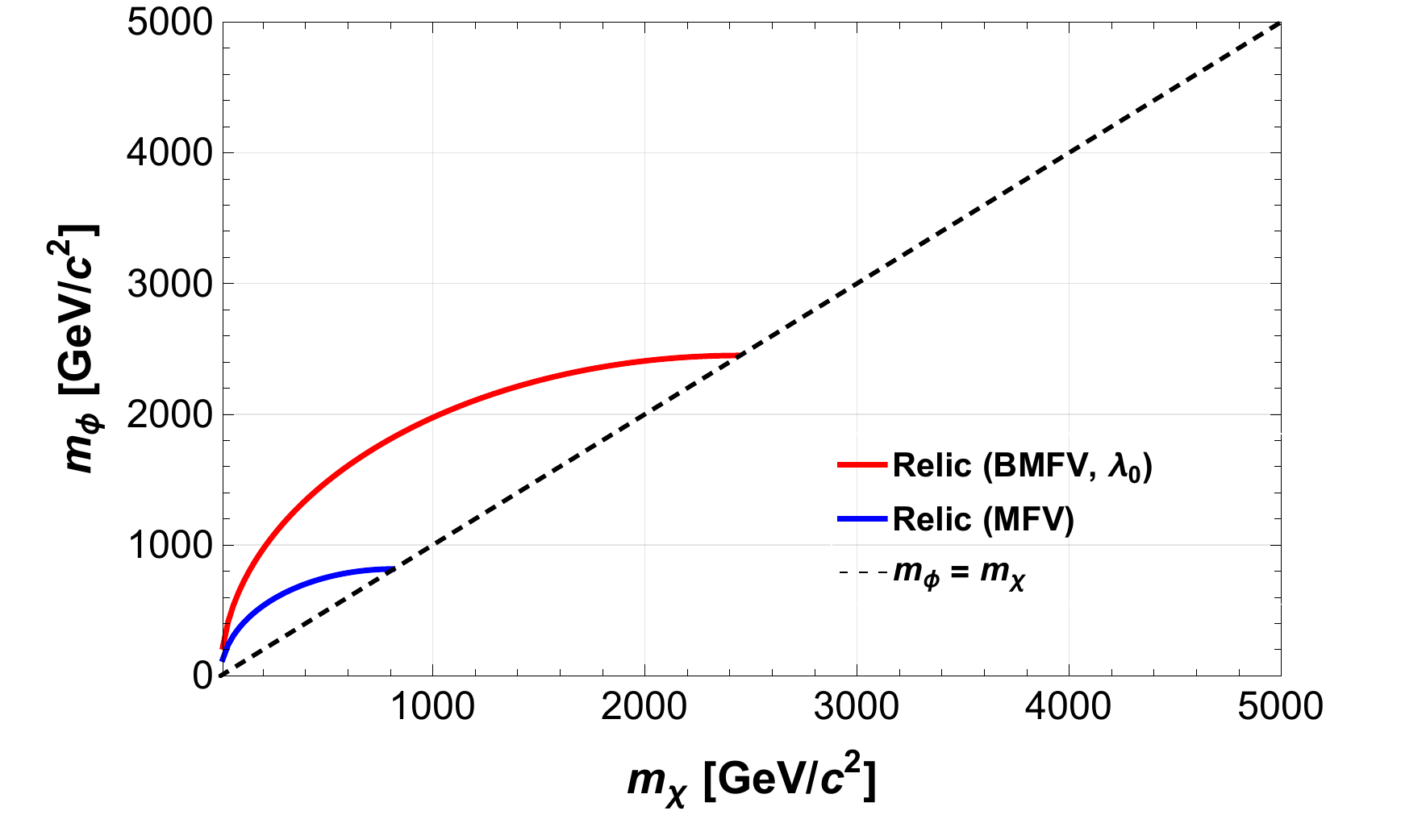}
\end{minipage}%
\begin{minipage}{.5\textwidth}
  \centering
  \includegraphics[width=1\linewidth]{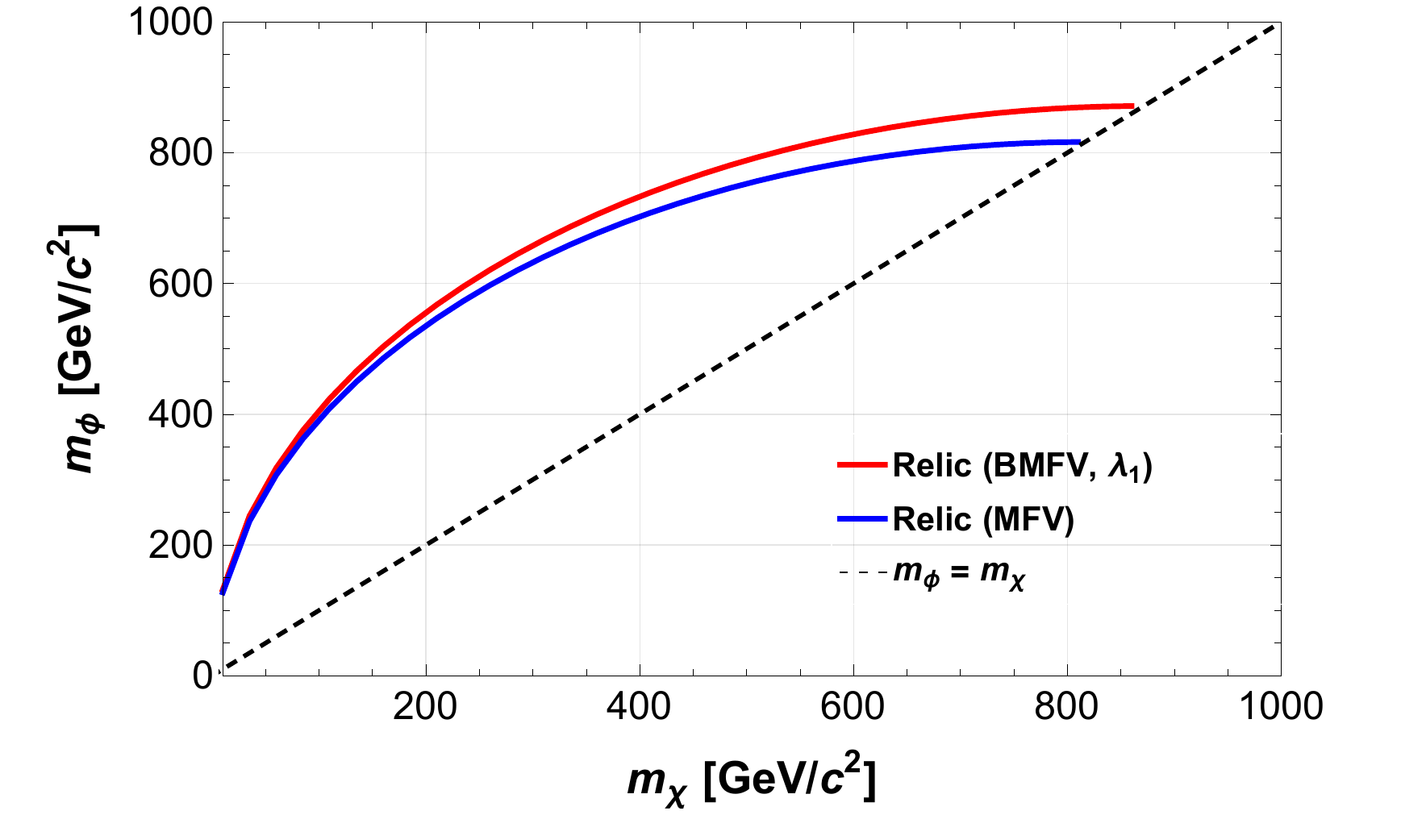}
\end{minipage}
\caption{The solid curves correspond to values of $m_{\phi}$ and $m_{\chi}$ that give rise to the correct relic abundance, 
for the choice of DM coupling matrix
$\lambda = \lambda_0$ (left)
and $\lambda = \lambda_1$ 
(right). For both cases, comparison with MFV is also shown.}
\label{fig:relic_param}
\end{figure}

On the other hand, in the case where the masses of the DM triplet components 
are highly degenerate, with $m_{\chi_a} \simeq m_{\chi_b} \simeq m_{\chi_c}$,
co-annihilation contributions
become important. This is the scenario assumed within
this work. If the species-transforming
processes $\chi_a l_i \rightarrow \chi_b l_j$ occur fast, all DM species 
are present at the freeze-out, and co-annihilation dominates. As described in the
Appendix \ref{app:calcrelic}, the effective 
annihilation cross-section,
to a good approximation \cite{Griest:1990kh}, 
will then be 
\begin{equation} \label{eq:relic_coann}
\langle \sigma v \rangle_{eff} =
\dfrac{1}{18} \sum_{i,j = e, \mu, \tau} \Big[\sum_{a,b = 1, 2, 3} 
\langle \sigma v \rangle_{\chi_a \overline{\chi}_b \rightarrow l_{i} \overline{l}_{j}}\Big] ,
\end{equation}
where the sum is performed over the $s$-wave contributions
of each channel.
This result is in agreement with \cite{Agrawal:2014aoa},
where quark-flavored DM with BMFV was considered. 
Equation~\eqref{eq:relic_coann} will apply for both BMFV and MFV.
The difference between the two scenarios will 
come from the additional BMFV contributions
in the sum, from the terms $\langle \sigma v \rangle_{\chi_a \overline{\chi}_b \rightarrow l_{i} \overline{l}_{j}}$
with $a \neq i, b \neq j$. 
With the effective cross-section taken to be
\begin{equation} \label{eq:effxsec}
\langle \sigma v \rangle_{eff} = 2.2 \times 10^{-26} \text{cm}^3/\text{s} ,
\end{equation}
which approximately gives the correct observed relic abundance \cite{Steigman:2012nb}, 
the $m_{\chi} - m_{\phi}$ parameter space can be constrained through 
Equation~\eqref{eq:relic_coann}. 
For simplicity, we have also assumed that the regime of interest
does not lie in the degenerate $m_{\chi} - m_{\phi}$ parameter space region,
where $\Delta_{\chi \phi} \equiv (m_{\phi} - m_{\chi}) / m_{\chi} \ll 1$ and that
$\Delta_{\chi \phi}$ is near or below
the freeze-out temperature $x_F$. 
To ensure DM stability, unless stated otherwise,
we have assumed that $m_{\phi} > m_{\chi}$.

In Figure~\ref{fig:relic_param}, we show  
the relic abundance constraints 
for the coupling matrix $\lambda$ choices of $\lambda_0$ and $\lambda_1$.
For a comparison, results for the case of MFV are also overlaid.
As expected, since $\lambda_1$ has a similar structure as the flavor-diagonal
MFV case, the relic abundance constraints are comparable between the two.
On the other hand, the $\lambda_0$ scenario shows that
relic abundance constraints are greater than those of MFV,
which is a direct consequence  
of the increased number of open 
channels in the BMFV case where all the flavor-violating $\lambda$ couplings significantly contribute.
We note that both MFV and BMFV cases  
considered above
are different from the setting of lepton MFV in \cite{Bai:2013iqa}, which focused on
the region of non-degenerate DM masses without the
co-annihilation effects and which considered the coupling to only one flavor at a time.

\subsection{Detection}
\label{sec:detection}

\subsubsection{Direct detection}
\label{sec:directdetection}

\begin{figure}[htb] 
    \centering
	\normalsize
\includegraphics[width=0.4\linewidth]{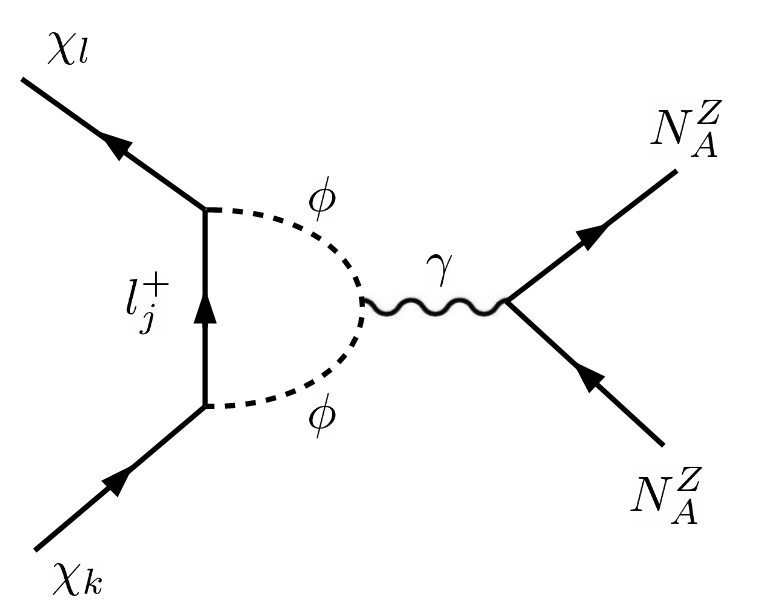}
\caption{Contributing diagram for the direct detection.}
\label{fig:directdetect_fig}
\end{figure}

In the direct detection experiments, dark matter interactions
with ordinary matter are studied. For the case of lepton-flavored dark matter,
unlike the case of quark-flavored DM \cite{Agrawal:2011ze, Bai:2013iqa},
there is no direct tree-level coupling to the target nucleus.
The only interactions which may occur for lepton-flavored DM
are the scattering from the target electron \cite{Kopp:2009et}
or through the photon exchange with the target nucleus which is a loop level process.  
As the scattering from the target electrons is highly suppressed,  
we will consider only the scattering off the target nucleus, which can occur in our model 
through the depicted diagram shown in 
Fig.~\ref{fig:directdetect_fig}. 

In principle, both spin-dependent (SD)  
and spin-independent (SI) 
DM interactions can 
contribute, through a dipole-dipole coupling and a charge-charge/charge-dipole coupling, respectively. 
For the lepton-flavored DM,
the SI charge-charge contribution dominates \cite{Agrawal:2011ze, Bai:2014osa}. 
The resulting 
DM-nucleon cross-section\footnote{The DM-nucleus cross-section $\sigma_{\chi}^\text{T}$ is found from
$\sigma_{\chi}^\text{T} = \sigma_{\chi}^\text{N} \cdot A^2$.} is
\begin{equation}
\sigma_{\chi}^{N} = \sum_{j = e, \mu, \tau} c_{j}^2 e^2 \Big(\dfrac{Z}{A} \Big)^2 \dfrac{\mu^2}{\pi} ,
\end{equation}
where $Z$ is the charge of the nucleus,
$A$ is the atomic number, $e$ is the electromagnetic coupling
and $\mu$ is the reduced 
DM-nucleon mass\footnote{Reduced mass is given by $\mu = m_{\chi} m_N / (m_{\chi} + m_N)$,
where $m_N$ is the mass of a nucleon.}.
The coefficients $c_{j}$ are given by \cite{Agrawal:2011ze}
\begin{equation}
c_{j} = - \frac{e}{ (64 \pi^2 m_{\phi}^2)}  \Big[\sum_{k, l = 1}^{3} (\lambda_{j k}^{\ast} \lambda_{j l}) \Big]
\Big[ 1 + 
\dfrac{2}{3} \log \Big( \dfrac{m_{l_j}^2}{m_{\phi}^2} \Big) \Big] ,
\end{equation}
with $m_{l_j}$ being the mass of the lepton $j$ running in the loop.
For the total cross-section we have summed over the possible
$e, \mu, \tau$ contributions. 
\begin{figure}[bp]
\centering
\begin{minipage}{.5\textwidth}
  \centering
  \includegraphics[width=1\linewidth]{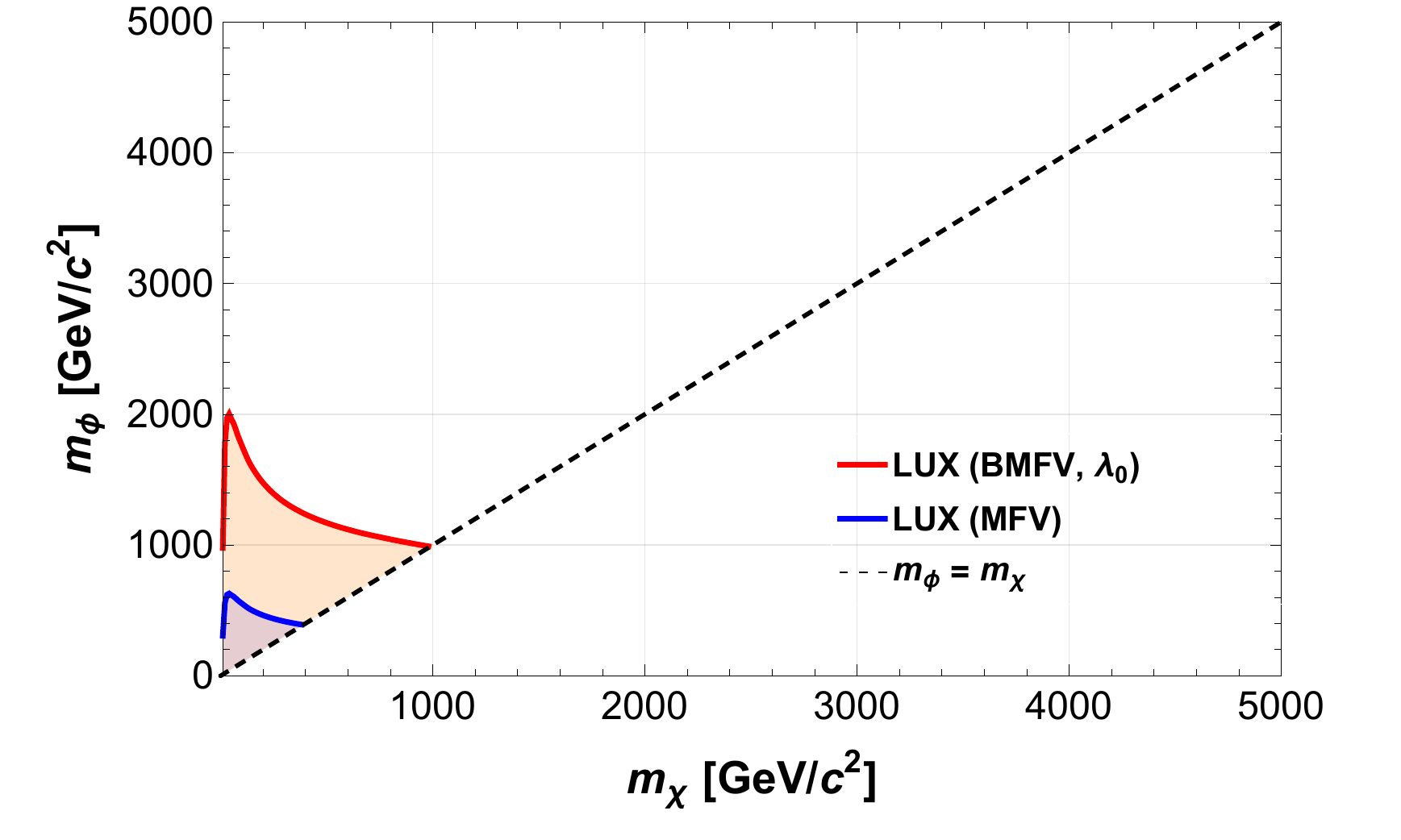}
\end{minipage}%
\begin{minipage}{.5\textwidth}
  \centering
  \includegraphics[width=1\linewidth]{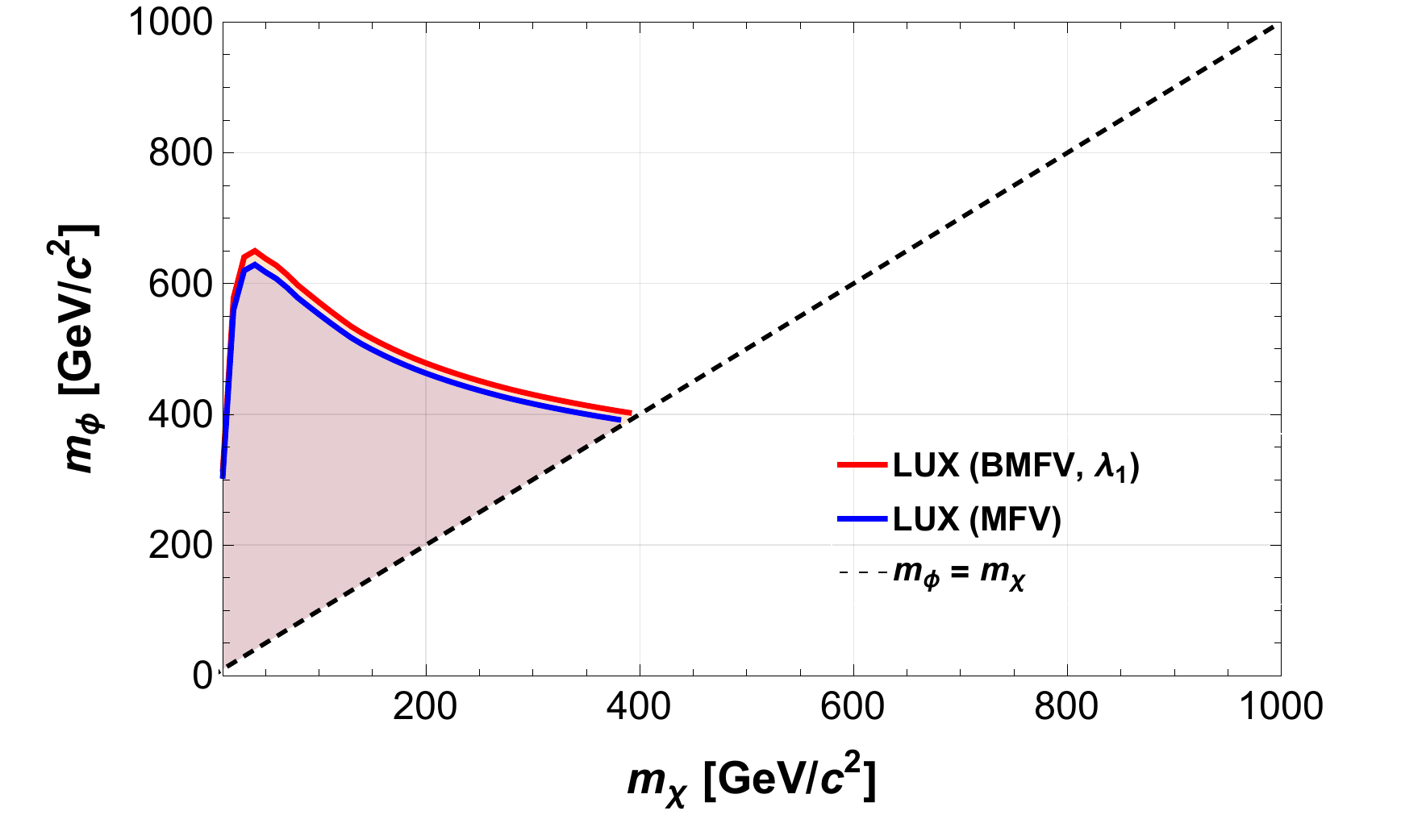}
\end{minipage}
\caption{Allowed parameter space  
by the direct detection constraints from LUX 
for the choice of DM coupling matrix
$\lambda = \lambda_0$ (left)
and $\lambda = \lambda_1$ 
(right).}
\label{fig:direct_param}
\end{figure}
Using the current result from LUX\footnote{For Xenon, $Z = 54$ and $A = 129$.}~\cite{Akerib:2013tjd},  
the constraints on the $m_{\phi} - m_{\chi}$ 
 parameter space is shown in Figure~\ref{fig:direct_param} 
for $\lambda = \lambda_{0}$ and $\lambda = \lambda_{1}$. 

\subsubsection{Indirect detection (A): electron--positron fluxes} 
\label{sec:indirectdetectionA}

Indirect detection constraints for lepton-flavored DM originate
predominantly from the electron-positron flux.
The AMS-02 measured \cite{Aguilar:2013qda} an  
excess in the position flux at high energies, which is 
difficult to explain from purely astrophysical sources
since they typically produce electron-dominant flux.
On the other hand, in DM scenarios such lepton-flavored DM with either 
MFV or BMFV assumption, electrons and positrons are produced in
equal amounts and thus could potentially explain the excess.
The main contribution to the AMS-02 spectrum will come from 
DM annihilation 
into positrons, $\chi_{a} \overline{\chi}_b \rightarrow e^{+} e^{-}$. The secondary positron
production from  $\mu^{+}$ or $\tau^{+}$ decays  
will result in a smeared
momentum distribution of the positrons and is thus weaker,
especially given that majority of this region is already
restricted from direct detection which 
is more sensitive than indirect detection.
Thus, we will only consider the channels 
with direct 
decays into $e^{+} e^{-}$.
\begin{figure}[htb]
\centering
\begin{minipage}{.5\textwidth}
  \centering
  \includegraphics[width=1\linewidth]{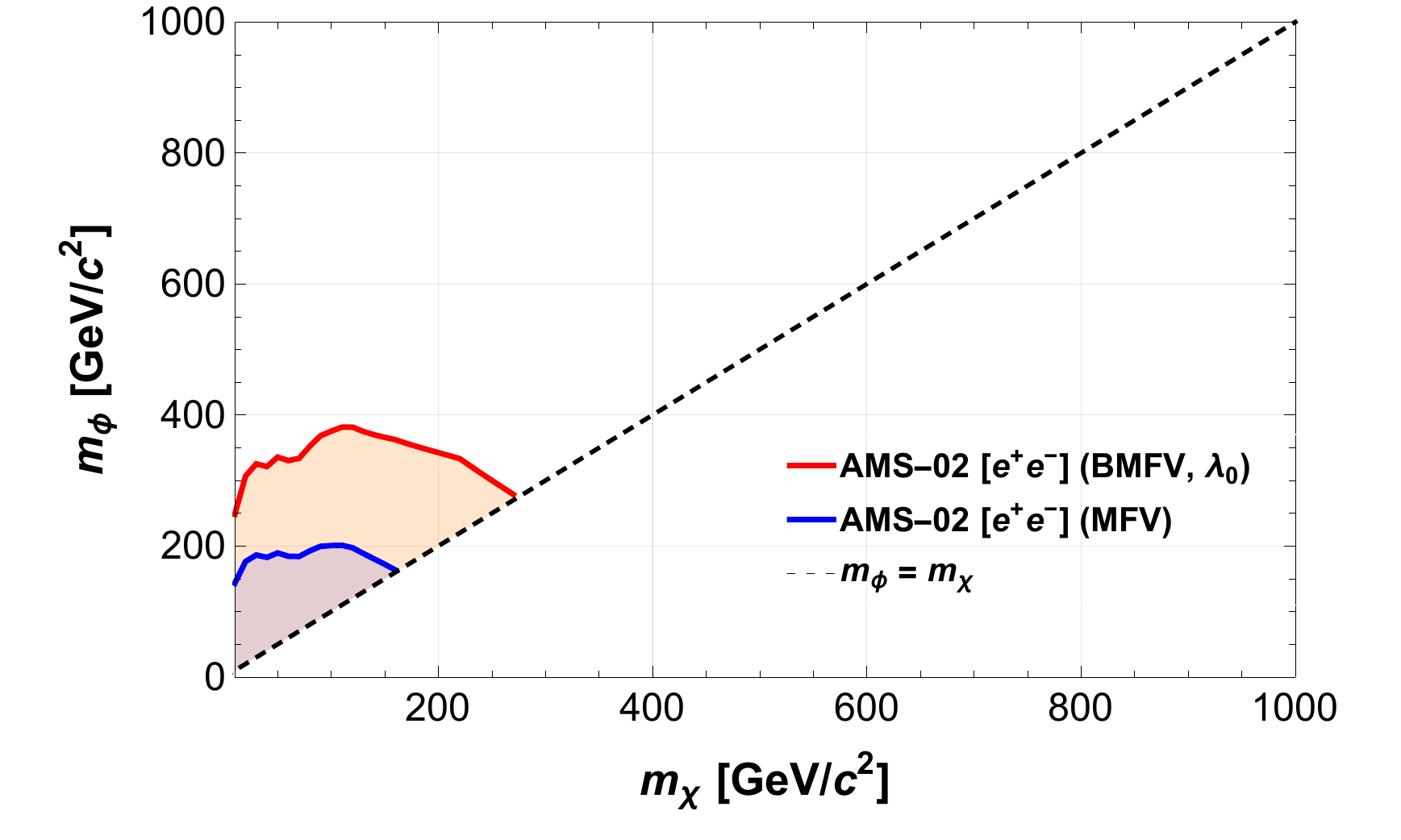}
\end{minipage}%
\begin{minipage}{.5\textwidth}
  \centering
  \includegraphics[width=1\linewidth]{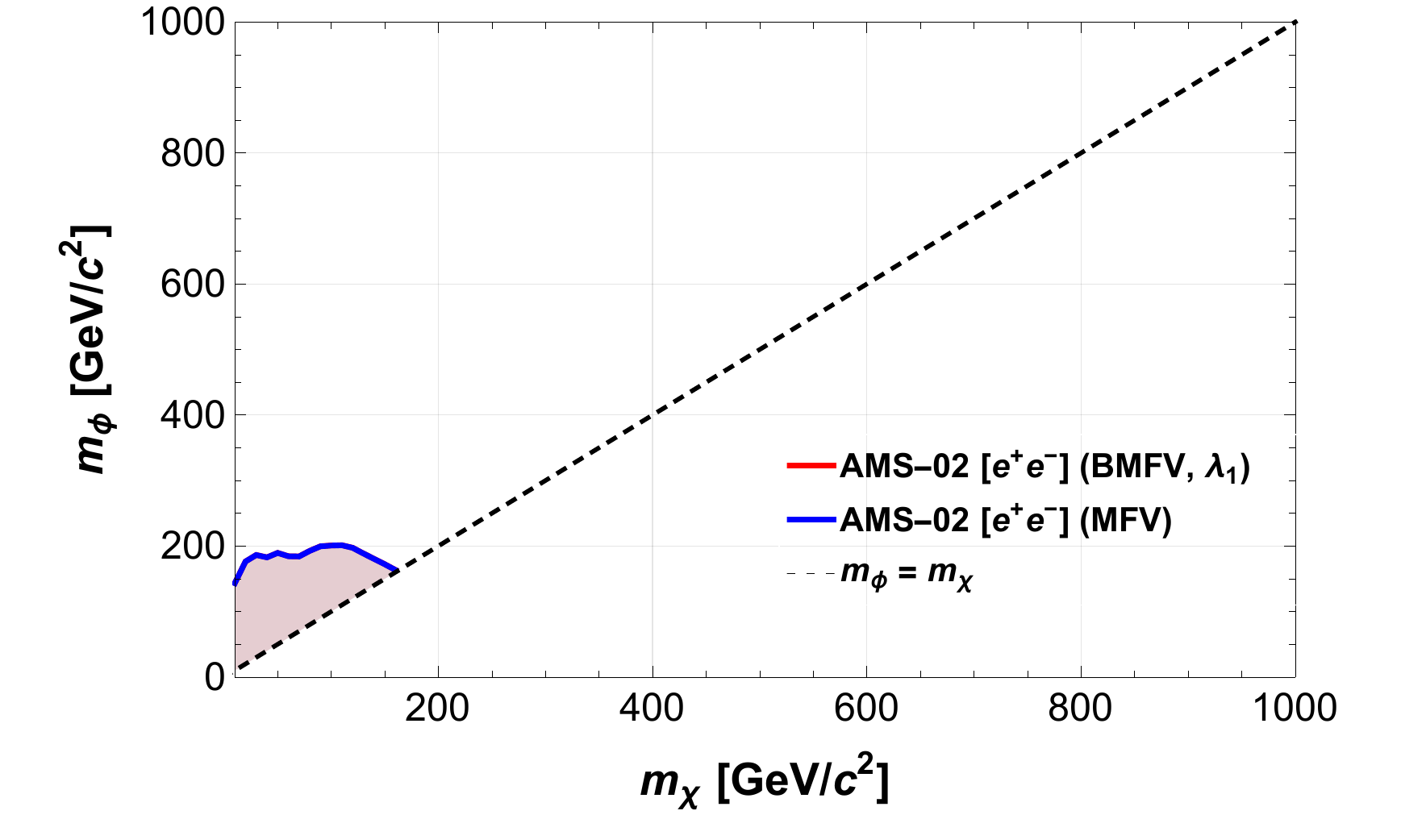}
\end{minipage}
\caption{Allowed parameter space  
by the indirect detection constraints 
from electron--positron fluxes 
for the choice of DM coupling matrix
$\lambda = \lambda_0$ (left)
and $\lambda = \lambda_1$ 
(right).}
\label{fig:indirect_param}
\end{figure}
Following \cite{Agrawal:2015tfa}, the $m_{\phi}-m_{\chi}$  
parameter space in our model with BMFV is constrained by the AMS-02 data 
similar to the way where the constraint on SUSY DM
decaying into $e^+ e^-$ is derived, after properly taking into account the factor of 2 difference 
between the Dirac (as in our model) vs Majorana (as in SUSY case) fermions in the annihilation cross section. 
The results are shown in Figure~\ref{fig:indirect_param},
for the two choices of coupling matrix $\lambda_0$ and $\lambda_1$. 
For this figure, we have used the $e^+e^-$ data of \cite{Ibarra:2013zia}
and considered only the $\chi_{a} \overline{\chi}_b \rightarrow e^+e^-$ channels.

\subsubsection{Indirect detection (B): $\gamma$ rays}
\label{sec:indirectdetectionA}

\begin{figure}[htb]
\centering
\begin{minipage}{.5\textwidth}
  \centering
  \includegraphics[width=1\linewidth]{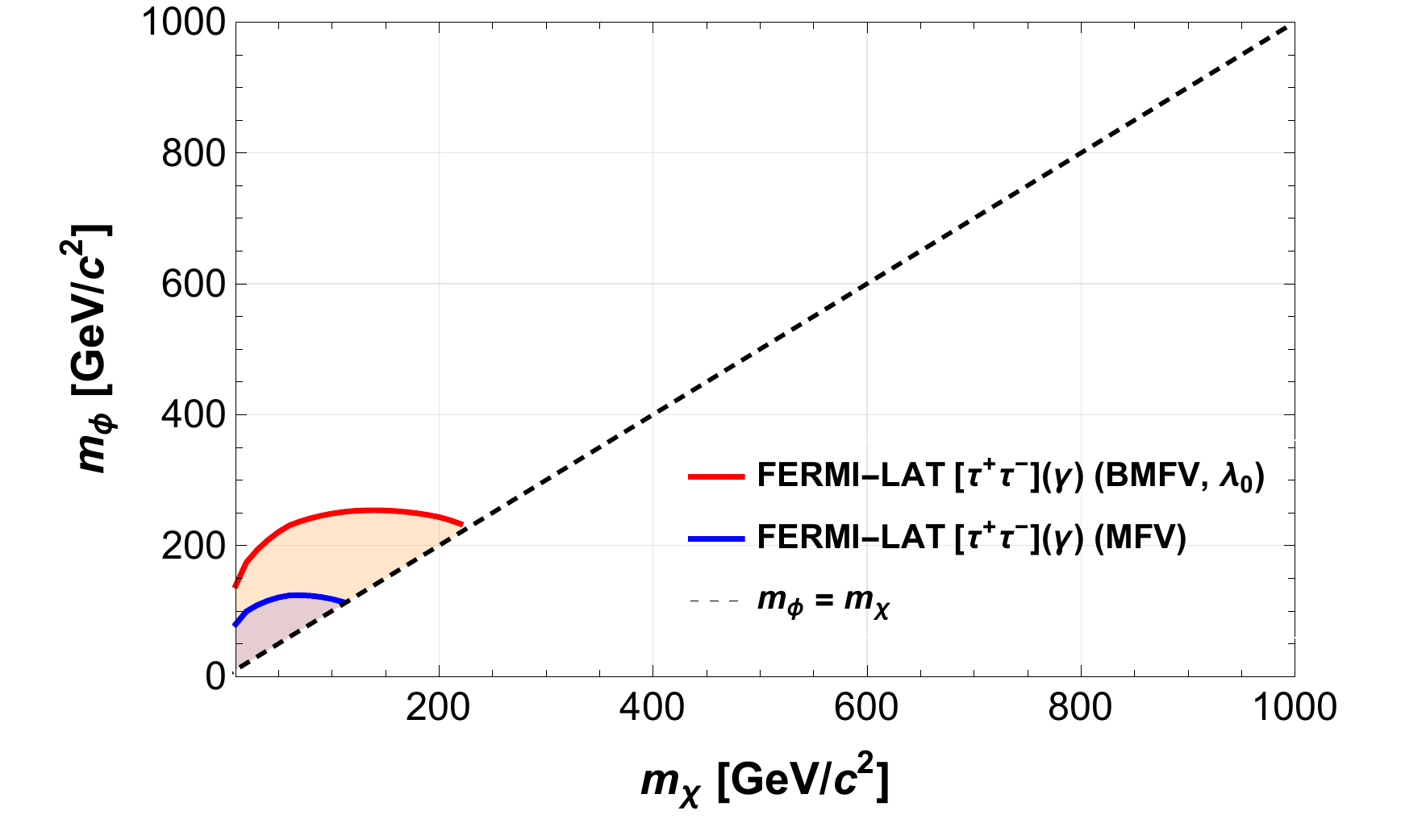}
\end{minipage}%
\begin{minipage}{.5\textwidth}
  \centering
  \includegraphics[width=1\linewidth]{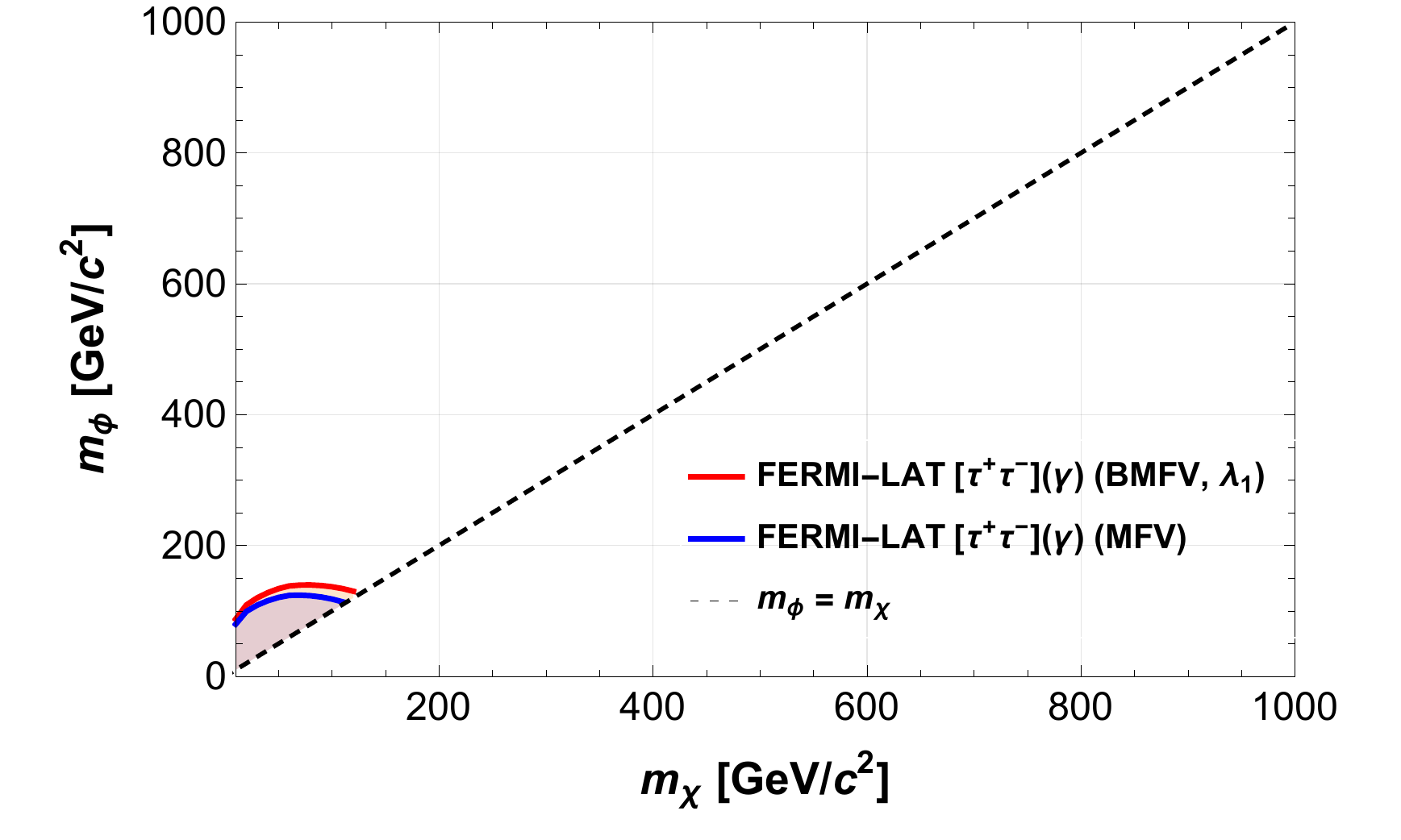}
\end{minipage}
\caption{Allowed parameter space 
by the indirect detection constraints
from $\gamma$-ray fluxes,  
 for the choice of DM coupling matrix
$\lambda = \lambda_0$ (left)
and $\lambda = \lambda_1$ 
(right).}
\label{fig:indirect_gamma_param}
\end{figure}

In addition to the position flux,   
the annihilation cross section is also bounded by the limits on the $\gamma$-ray sources. 
The $\gamma$-rays can arise from various processes. These include 
the annihilation channel through lepton--anti-lepton intermediate state, 
$\chi_{a} \overline{\chi}_{b} \rightarrow l^{+}_{i} l^{-}_{j}
\rightarrow \gamma$. 
The production of photons can also arise through the hadronic $\tau$ decay channels 
such as $\tau \rightarrow \pi^+ \pi^0 \nu$\footnote{This is the main $\tau$ decay channel,
with a branching ratio of 25.52\%.}, followed
by $\pi^0 \rightarrow \gamma \gamma$, which turns out to be the dominant contribution.  
In Figure~\ref{fig:indirect_gamma_param}, 
we present the constraints from the $\gamma$-ray
flux on the 
$\chi_{a} \overline{\chi}_b \rightarrow \tau^+ \tau^-$ channel,
which are based on simulations \cite{Tavakoli:2013zva}
consistent with Fermi-LAT data.
We show results for both $\lambda_0$ and $\lambda_1$, which turns out to be 
less stringent than those from the direct detection.

While the $\gamma$-ray constraints are not overly stringent,  
there have been several (potential) observations of 
$\gamma$-ray line excesses which may be interpreted
in the context of DM. Many proposals 
have been put forward as a potential explanation for the 
recently observed galactic 3.5 keV X-ray excess \cite{Bulbul:2014sua, Boyarsky:2014jta}.
Within the lepton-flavored dark matter framework~\cite{Agrawal:2015tfa}, one possibility  
is to consider 
the energy release
after one DM species decays 
to another through
a flavor changing process $\chi_{a} \rightarrow \chi_{b} \gamma$ at one loop, with the 
energy of $\gamma$ identified with the mass splitting of the species
$\Delta{m_{ab}} = m_{a} - m_{b}$. Another possibility is the 
decay $\chi_{a} \rightarrow \chi_{b} l_i \overline{l}_j$
due to flavor preserving couplings, 
where the lepton is identified with a neutrino if the mass splitting $\Delta m_{ab}$  
is smaller than $2 m_e$. 
Generally, the former dominates over the latter, given that the latter case is further suppressed by the 3-body decay phase space as well as additional powers of loop lepton masses. 
The mass splitting required to explain the 3.5 keV $\gamma$-ray line 
can be generated within the lepton-flavored DM framework with MFV~\cite{Agrawal:2015tfa}. 
At the leading order the $\chi_{a}$ masses are degenerate due to the flavor symmetry both in the MFV and BMFV cases. 
On the other hand, the splitting among the $\chi_{a}$ masses is induced 
at the one-loop level through wave-function renormalization, which will depend on the loop charged fermion masses. 
This will induce several $\gamma$-ray lines with closely spaced 
frequencies
proportional to mass splitting, given by
\begin{equation} \label{eq:linesplit}
 \delta w = w (\chi_{\tau} \rightarrow \chi_{e}) - w (\chi_{\tau} \rightarrow \chi_{\mu}) \approx 
w_0 (m_{\mu}^2 /m_{\tau}^2) ,
\end{equation}
where $w (\chi_{\tau} \rightarrow \chi_{e})$ and $w (\chi_{\tau} \rightarrow \chi_{\mu})$ are the frequencies 
for the transitions $\chi_{\tau} \rightarrow \chi_{\mu}$ and $\chi_{\tau} \rightarrow \chi_{\mu}$, respectively. 
The parameter $w_0$ is the average $\chi_{\tau}$ line frequency,
 which if taken to be 3.5 keV to explain the excess will result in $\delta w \simeq 12.4$ eV.
Hence, not only will the excess be explained but there will be 
an eV level line splitting at 3.5 keV.

Similarly, we can use the same argument to address the 
long standing 511 keV $\gamma$-ray excess observed by SPI/INTEGRAL 
\cite{Knodlseder:2003sv,Jean:2005af} in the Galactic Center.
By setting $w_0 = 511$ keV in \eqref{eq:linesplit},
there will be an induced splitting $\delta w \simeq 1.82$ keV, 
which is still consistent with the observed broadening of several keV
around the 511 keV line \cite{Jean:2005af} and will require future experiments
to settle the question if such features are indeed present.

\subsection{Collider searches}
\label{sec:collider}

We will now comment on hadron ($pp$) and lepton ($e^+ e^-$) collider searches,
emphasizing the difference between MFV and BMFV scenarios.
For our studies, we have estimated cross-sections at the parton level using
\texttt{MadGraph5}~\cite{Alwall:2011uj}, with the flavored DM models implemented 
via \texttt{FeynRules}~\cite{Alloul:2013bka}.~Since 
the final states of interest are associated with leptons,
hadronization and associated showers can be neglected.
We will mention possible backgrounds only in passing, with a
detailed background estimates and analysis for various scenarios left for future work.
In the simulations, unless stated otherwise, we have used the extreme case of $\lambda_0$ 
to represent BMFV, in order to highlight the difference with MFV.

\subsubsection{Hadron colliders}
\label{sec:hadroncollider}

\begin{figure}[htb]
\centering
\begin{minipage}{.5\textwidth}
  \centering
  \includegraphics[width=0.5\linewidth]{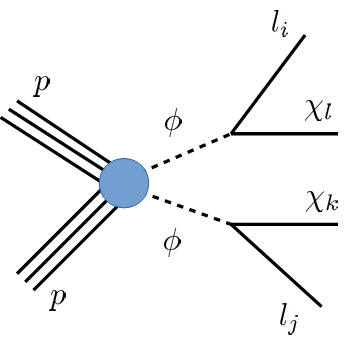}
\end{minipage}%
\begin{minipage}{.5\textwidth}
  \centering
  \includegraphics[width=1\linewidth]{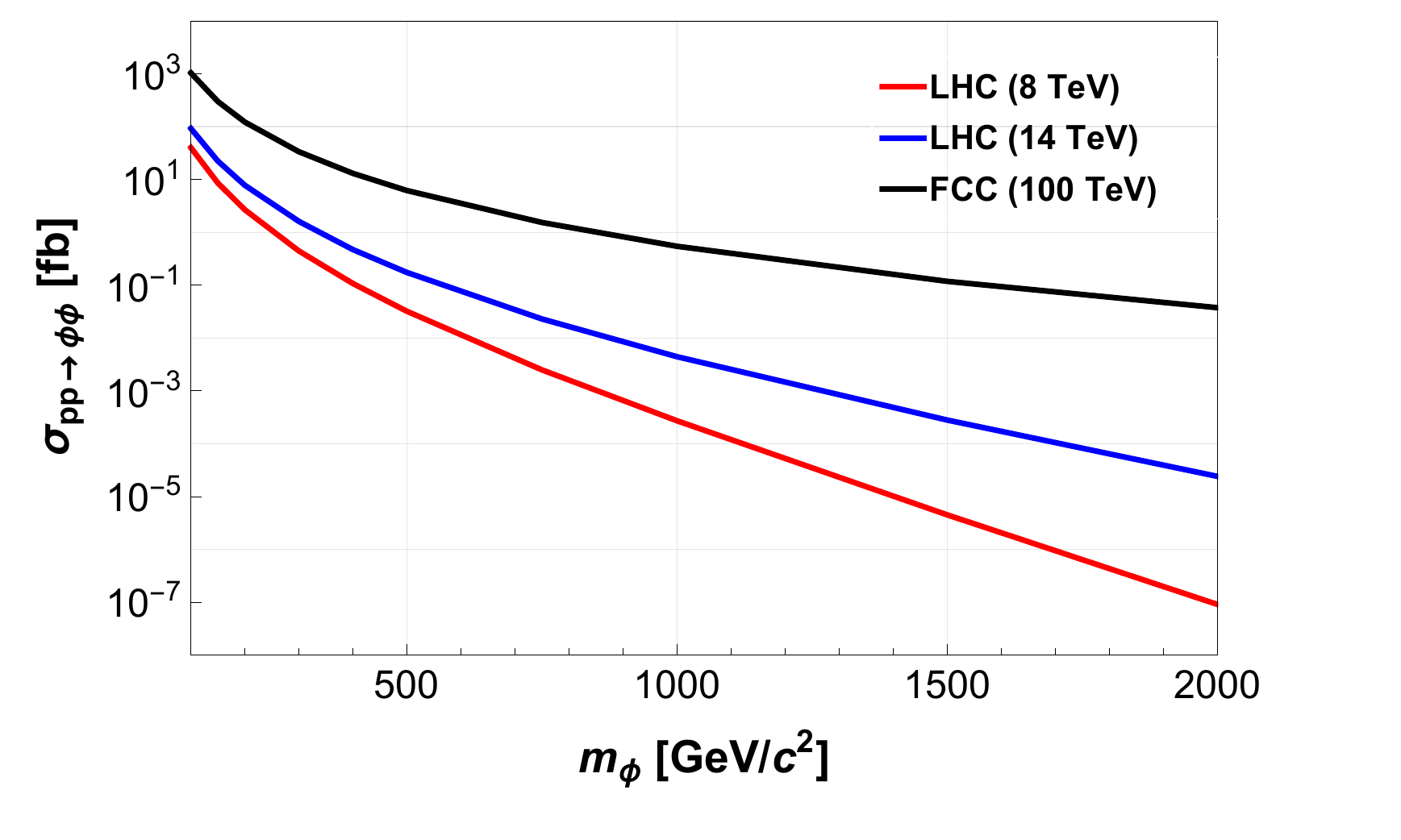}
\end{minipage}
\caption{(left) Mediator pair production at hadron colliders.
(right) Production cross-section
as a function of the mediator mass for LHC as well as FCC.}
\label{fig:hadron_medproduction}
\end{figure}

The most sensitive hadron collider signals for the lepton-flavored DM
come from Drell-Yan production of the mediator pair through
$Z$ and $\gamma$ with $\chi \overline{\chi} l^{+} l^{-}$ final state,
giving a signature of same flavor, opposite sign di-lepton plus missing energy ($l^{+} l^{-}$ + MET).
 The general sensitivity of these searches
can be seen from the mediator production cross-section, 
as shown in Figure~\ref{fig:hadron_medproduction}, 
which is the same for both MFV and BMFV. 
The results for LHC ($\sqrt{s} = 8$ and $14$ TeV), in the mass region below
1 TeV, are in agreement with \cite{Agrawal:2011ze, Bai:2014osa}.
For potential future studies, we have extended the considered mass region
and also shown the results for the proposed far
future circular collider (FCC) operating at $\sqrt{s} = 100$ TeV.

\begin{figure}[htb]
\centering
  \includegraphics[width=0.5\linewidth]{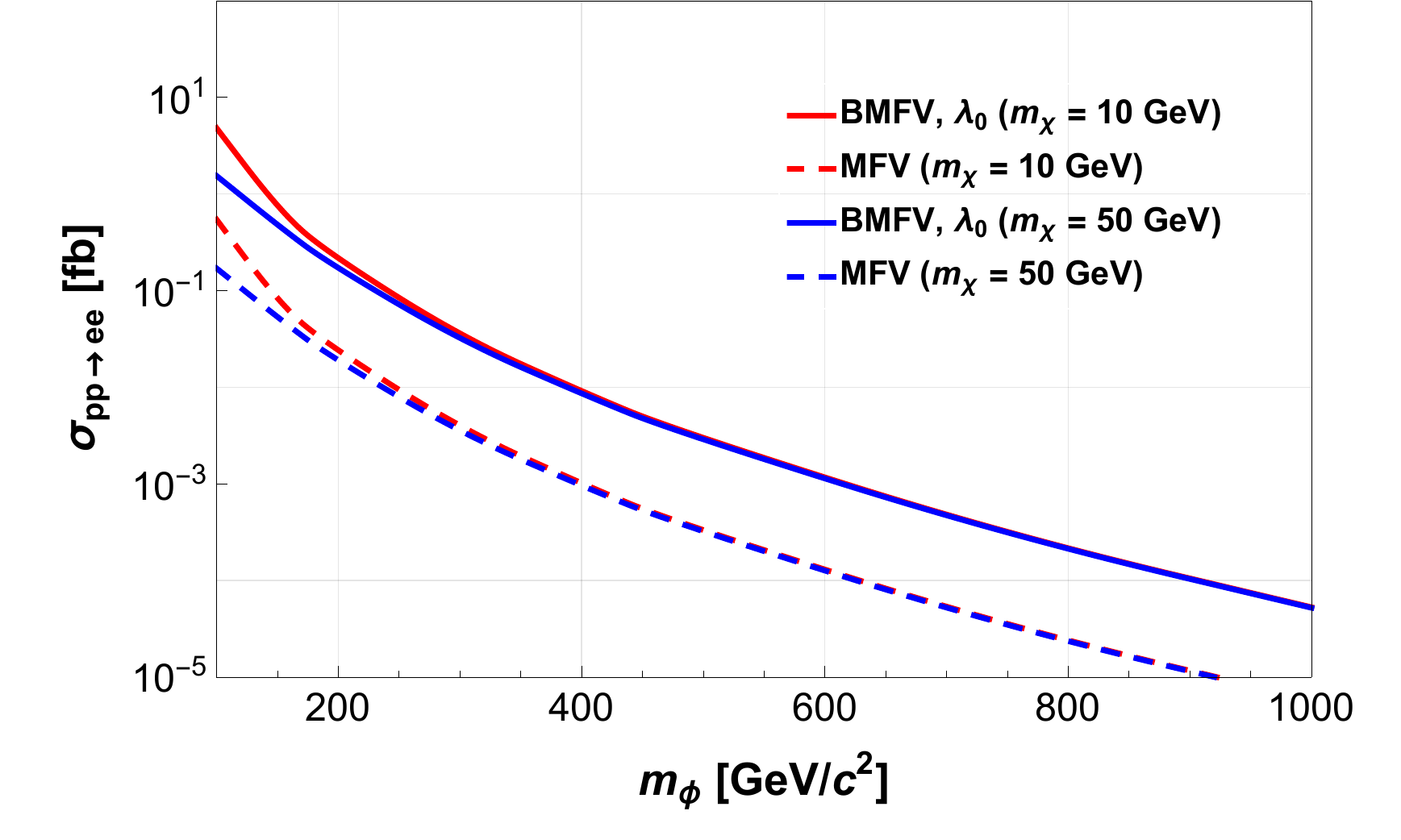}
\caption{Comparison of MFV and BMFV cross sections for $pp \rightarrow e^+ e^-$ production
at the LHC (14 TeV), assuming $m_{\chi} = 10$ GeV and $m_{\chi} = 50$ GeV.}
\label{fig:lhc_eeproduction}
\end{figure}

In general, BMFV will have enhanced di-lepton signal over that in the MFV case, 
due to the existence of additional contributing $\chi$ channels.
Note that $l^{+} l^{-} +$ MET is also the standard signature for SUSY slepton searches. 
Hence, based on slepton search analyses from ATLAS \cite{Aad:2014vma} and CMS \cite{Khachatryan:2014qwa}, 
the constraints on the flavored DM model can be obtained \cite{Agrawal:2011ze, Bai:2014osa,Chang:2014tea}. 
As an illustration, in Figure~\ref{fig:lhc_eeproduction} we display both MFV and BMFV
cases for the simplest final state of $e^+ e^-$ at LHC (14 TeV),
assuming two representative scenarios of $m_{\chi} = 10$ GeV and $m_{\chi} = 50$ GeV.
We anticipate that the upcoming LHC 14 TeV run will provide additional
constrains on the parameter space in this region.
Results for other lepton final states are similar. 

On the other hand, the multi-flavor lepton final state (e.g. $e^{\pm}\mu^{\mp}$) 
in SUSY result from chargino production. Since the event topologies for the
chargino production and flavored-DM differ, the two cannot be mapped directly to each other.
The main background in these searches comes from di-boson ($WW, ZW, ZZ$)
as well as top production. Additional signatures could come from taus in the
final state, and these would be the main collider search channels which could 
discriminate between BMFV (with $\lambda_1$ or $\lambda_{2}$) and MFV, due to large 
off-diagonal couplings involving $\tau$. These channels, however,
are less sensitive than the others, due to high background contamination
and lower efficiency \cite{Aad:2014mra}.

\subsubsection{Lepton colliders}
\label{sec:leptoncollider}

\begin{figure}[htb]
\centering
\begin{minipage}{.5\textwidth}
  \centering
  \includegraphics[width=0.5\linewidth]{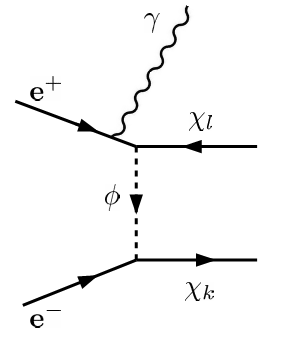}
\end{minipage}%
\begin{minipage}{.5\textwidth}
  \centering
  \includegraphics[width=1\linewidth]{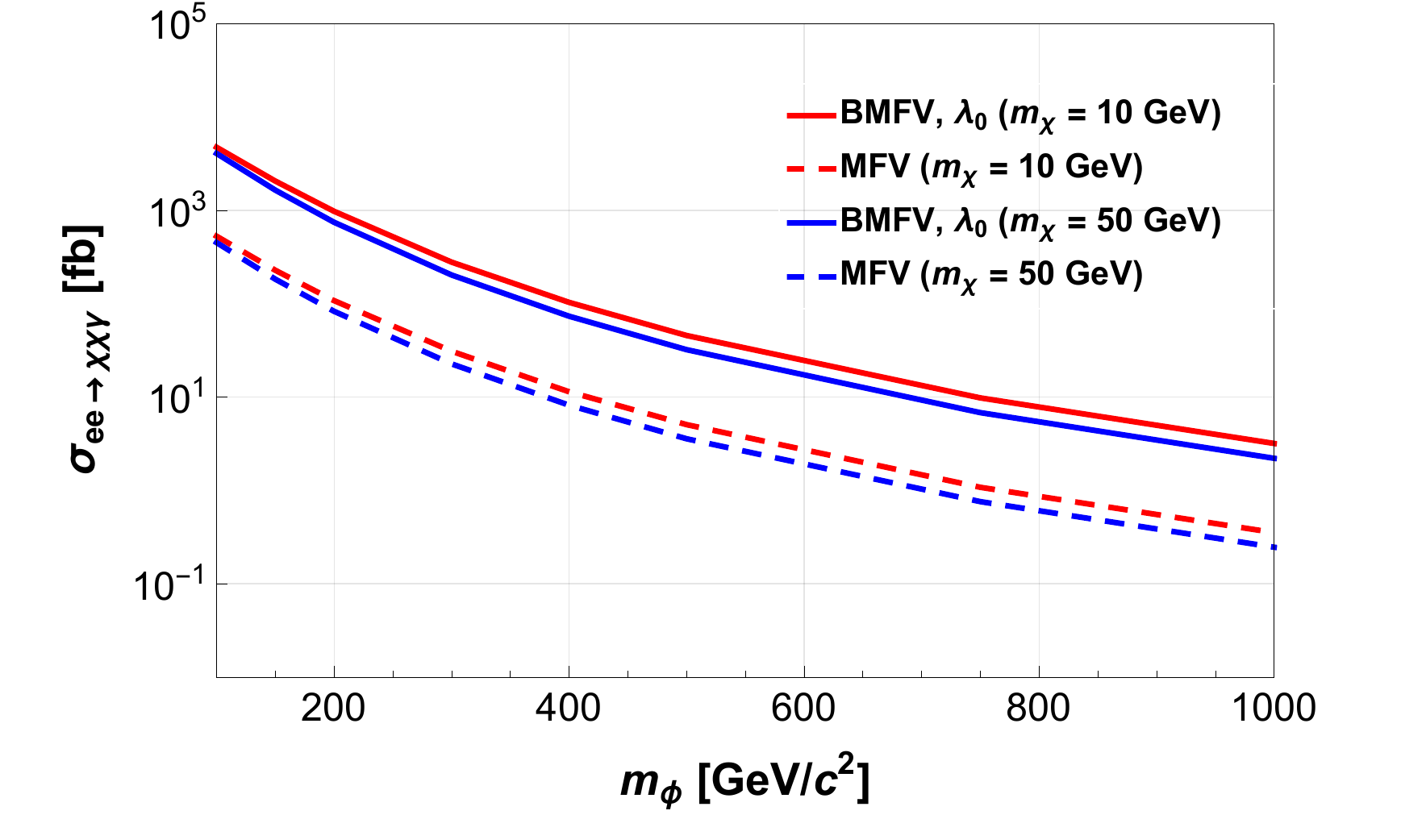}
\end{minipage}
\caption{(left) Contributing diagram for $e^+ e^-$ collider 
mono-photon channel. (right) LEP mono-photon channel cross-section.}
\label{fig:lep_monophoton}
\end{figure}

For $e^+e^-$ lepton collider at LEP ($\sqrt{s} = 200$ GeV),  
one of the best discovery channels is
$e^+e^- \rightarrow \chi \overline{\chi} \gamma$ with search signature of a mono-photon 
plus missing energy ($\gamma +$ MET).
A possible contributing diagram 
as well as general parton-level cross sections
are shown in Figure~\ref{fig:lep_monophoton}.
While proper comparison without detailed analysis is difficult, 
from Fig. \ref{fig:lhc_eeproduction} and \ref{fig:lep_monophoton}, 
it is evident that constraints from lepton collider are more significant. 
In fact, it has been pointed out that using effective field theory (EFT)
approach \cite{Fox:2011fx}, in the context of lepto-philic DM
with a charged scalar mediator, similar to MFV and BMFV models
we consider, the LEP DELPHI experiment results~\cite{Abdallah:2003np} 
restrict the DM mass to be $\gtrsim 100$ GeV and mediator 
mass to be above several hundred GeV.
The dominant background for this search,
is the $e^+e^- \rightarrow Z \gamma$ process,
with $Z$ decaying invisibly via $Z \rightarrow \nu \overline{\nu}$.

\begin{figure}[htb]
\centering
  \includegraphics[width=0.25\linewidth]{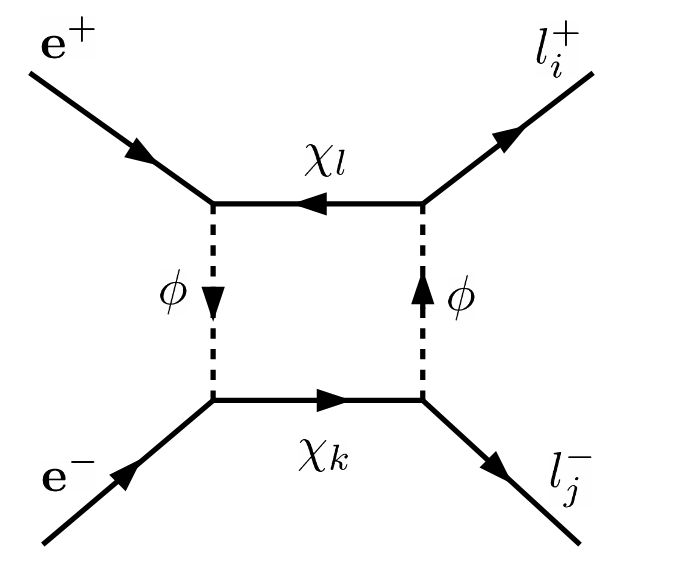}
\caption{Contributing diagram for $e^+ e^-$ collider 
fermion pair production channel.}
\label{fig:lep_pair_production}
\end{figure}

Aside from the mono-photon searches, for the case of lepton-flavored DM with
a charged scalar mediator, the fermion pair production is important.
A diagram that can lead to fermion pair production is shown in Figure~\ref{fig:lep_pair_production}.
In fact, recent EFT analysis \cite{Freitas:2014jla} shows that LEP fermion 
pair production constraints \cite{Abreu:2000ap} are competitive 
with mono-photon searches and limit the mediator mass in the region of 
several hundred GeV and DM mass below 100 - 200 GeV. 
Additional advantage of searching in these channels is that 
it provides a way to test the flavor violating couplings, which is not possible in  
the mono-photon channel as this channel only probes DM coupling to electrons. 
In BMFV one can expect processes with multi-flavor lepton final state
 such as $e^+e^- \rightarrow \mu^+e^-$, with
$\chi$ and $\phi$ running in the loop.
The scenario can thus be tested by searching for multi-flavor lepton final state.
Past experimental searches mainly focused on single flavor lepton final state 
and thus these channels are not as constrained.

Looking into the future, the proposed International Linear Collider (ILC)
with the collision energy of $\sqrt{s} = 500$ GeV is expected 
to probe \cite{Freitas:2014jla} the TeV region in $m_{\phi}, m_{\chi}$ parameter space, thus
putting stringent bounds on the model.


\subsection{Decaying dark matter}
\label{ssec:dmdecay}

Having investigated the constraints and signals
for the case of stable dark matter, we will also briefly discuss
decaying dark matter. As we have shown in 
Section~\ref{sec:dmstability}, there is no ``natural'' stabilizing
symmetry for dark matter present within the lepton-flavored MFV or BMFV scenarios.
Hence, if one is to ensure DM stability, an extra symmetry such
as the $\Z2$ symmetry mentioned in Section~\ref{sec:dmstability} 
must be imposed ad hoc. On the other hand, 
without such stabilizing symmetry, our framework provides  
an interesting setup for studying decaying dark matter.
As has been stressed in the recent literature,
decaying dark matter can lead to novel experimental signatures,
especially for indirect detection \cite{Ibarra:2013cra}.

To obtain effective operators for decaying DM, we have scanned
for gauge invariant terms which obey
the constraints of equations from Section~\ref{sec:dmstability},
using a custom \texttt{Mathematica} code. 
The smallest operators which contain a single DM $\chi$ field ($N_{\chi} = 1$)
are 4-fermion dimension-6 terms with one lepton.
We found four combinations of such operators, corresponding to decays
\begin{equation}
\chi \rightarrow \text{lepton} + \text{meson}
\end{equation}
to which many combinations of couplings can be added to make
them invariant. Assuming a minimal combination
of couplings, these are:
\begin{align}
\Big(\dfrac{\lambda Y_D Y_U^{\dagger}}{\Lambda^2} \Big) \chi \overline{e_R} d_R \overline{u_R} 
&~~~ \Longrightarrow~~~  \chi \rightarrow e^+ \pi^- \\
\Big(\dfrac{\lambda Y_E Y_D}{\Lambda^2} \Big) \chi d_R \overline{L} \overline{Q}
= \Big(\dfrac{\lambda Y_E Y_D}{\Lambda^2} \Big) \chi d_R 
\{\overline{\nu_L} \overline{d_L} - \overline{e_L} \overline{u_L}\} 
&~~~ \Longrightarrow~~~  \chi \rightarrow \overline{\nu} \pi^0~,~\chi \rightarrow e^+ \pi^- \\
\Big(\dfrac{\lambda Y_E Y_U^{\dagger} }{\Lambda^2} \Big) \chi \overline{u_R} \overline{L} Q
= \Big(\dfrac{\lambda Y_E Y_U^{\dagger} }{\Lambda^2} \Big) \chi \overline{u_R} 
\{\overline{\nu_L} d_L - \overline{e_L} u_L\} 
&~~~ \Longrightarrow~~~  \chi \rightarrow \overline{\nu} \pi^0~,~\chi \rightarrow e^+ \pi^- \\
\Big(\dfrac{\lambda Y_E^{\dagger}}{\Lambda^2} \Big) \chi e_R \overline{L} \overline{L}
= \Big(\dfrac{\lambda Y_E^{\dagger}}{\Lambda^2} \Big) \chi e_R \{\overline{\nu_L} \overline{e_L} - \overline{e_L} \overline{\nu_L}\} 
&~~~ \Longrightarrow~~~  \chi \rightarrow \overline{\nu} e^+e^- 
\end{align}
where we have suppressed the flavor indices and 
taking for illustrative purposes first generation particles,
displayed the possible processes on the right. 
Since there exists a similarity between our DM $\chi$
and SUSY neutralino, the channels which appear above are
analogous to those of decaying neutralino \cite{Baltz:1997ar}
in $R$-parity violating (RPV) MSSM.
Namely, with lepton number ($L$) violating RPV operators $L Q \overline{d}$ and $LL\overline{e}$,
neutralino could also decay via $\chi \rightarrow \overline{\nu} d \overline{d}, 
\chi \rightarrow \overline{e} d \overline{u}$ and $\chi \rightarrow \overline{\nu}e \overline{e}$.

For the decaying DM to be consistent with observation, 
its lifetime must be longer than the age of the Universe, 
$\tau_{univ.} \sim 4.3 \times 10^{17}$s.
For the above 4-fermion operators, the lifetime is given by \cite{Eichler:1989br}
\begin{equation}
\tau_{\chi} \sim 10^{26}\text{s} \Big(\dfrac{1}{f(\lambda, \lambda^{\dagger}, Y, Y^{\dagger})}
\Big)^2\Big(\dfrac{\text{TeV}}{m_{\chi}}\Big)^5
 \Big(\dfrac{\Lambda}{10^{15}~\text{GeV}}\Big)^4
\end{equation}
where we have included the couplings using
 a function $f(\lambda, \lambda^{\dagger}, Y, Y^{\dagger})$, which denotes the 
appropriate combination of $\lambda$'s and Yukawa's
such that a given operator is rendered invariant.

In general, decaying dark matter can lead to 
indirect detection signals coming from gamma rays \cite{Dugger:2010ys,Hutsi:2010ai,Cirelli:2009dv,Ibarra:2009nw},
neutrinos \cite{Buckley:2009kw,Spolyar:2009kx,Covi:2009xn,Hisano:2008ah}, 
electrons/positrons \cite{Cirelli:2008pk,Nardi:2008ix,Ibarra:2008jk,Ibarra:2009dr,Ishiwata:2009vx}
and anti-protons/anti-deuterons \cite{Cui:2010ud,Kadastik:2009ts,Ibarra:2009tn,Evoli:2011id,Cholis:2010xb}.
From the above channels, if $m_{\chi}$ is not very high
($m_{\chi} \sim$ GeV), the $\chi$ decay will result in
a prompt hard lepton as well as several softer
leptons or gammas coming from the meson decay. If $m_{\chi} \gg$ GeV,
energetic quarks will lead to hadronizing jets resulting in
a platitude of even softer leptons or gammas\footnote{In this case, there will also be additional
constraints from astrophysics, such as those from
the anti-proton flux. We would like to thank Tim Tait for stressing this point.}.
Decaying DM with a hard charged lepton
can in principle be used to explain 
(for example, see \cite{Ibarra:2013cra,Hamaguchi:2008ta}) the positron flux 
excess observed by Pamela \cite{Adriani:2008zr} and AMS-02 \cite{Aguilar:2013qda}.
On the other hand, very heavy ($m_{\chi} \sim 100$ TeV)
DM with a prompt energetic neutrino, which can also appear 
within our setup, can have implications
for neutrino observatories such as IceCube or Super-Kamiokande \cite{Rott:2014kfa}.
Additionally, gamma ray signals have already been extensively analyzed
by the Fermi LAT \cite{Ackermann:2015lka} and HESS \cite{Abramowski:2013ax} experiments. 
Future experiments,
such as CTA \cite{Consortium:2010bc}, will allow to further investigate the
parameter space for heavier DM. Analyses of decaying DM with such signatures 
already exist in the literature \cite{Arcadi:2013aba,Garny:2012vt,Garny:2010eg}.
The constraints that we have obtained
 from the previous sections,
especially those from flavor, restrict the allowed structure in the
coupling $\lambda$. Hence,
an astrophysical analysis will result in the allowed
parameter space for $m_{\chi}$ and $\Lambda$.
A detailed study of decaying dark matter
and the implications of astrophysics, however, is
 beyond the scope of the present work and is left for future analysis.


\section{Summary}
\label{sec:summary}

\begin{figure}[htb]
  \centering
  \includegraphics[width=0.75\linewidth]{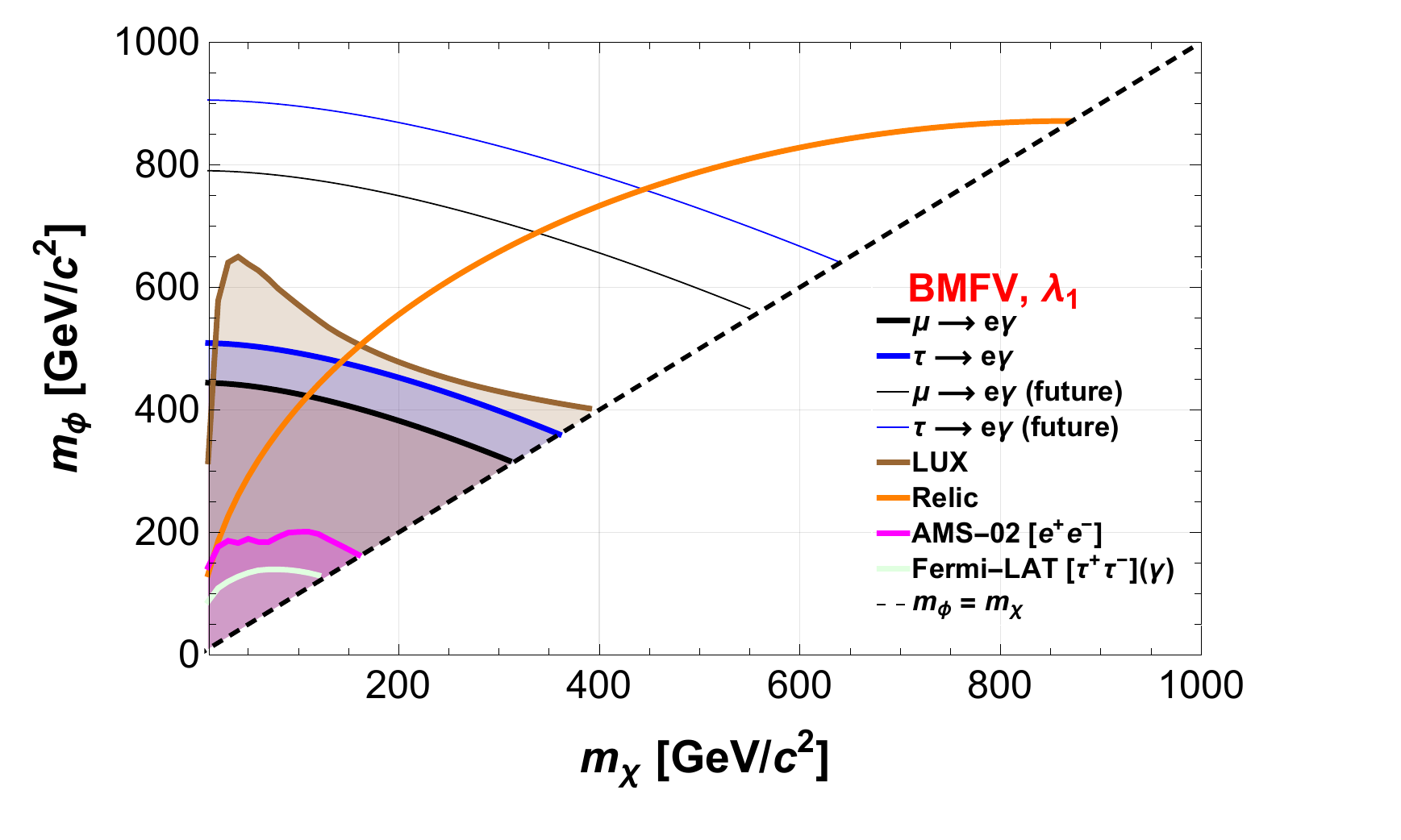}
\caption{Allowed parameter space for the coupling choice of $\lambda_1$,
after the constraints from LFV, relic abundance, direct and indirect
detection have been combined.}
\label{fig:summary_figure}
\end{figure}

In summary, BMFV lepton-flavored DM which transforms under additional
flavor $\SU3_{\chi}$ symmetry, provides an interesting possibility to explore
and contrast with MFV. In the literature it has been pointed out
that for quark-flavored MFV and BMFV scenarios there is a readily available
$\Z3$ symmetry which stabilizes the dark matter. We have shown,
that for lepton-flavored MFV and BMFV scenarios this is not the case,
and one needs to impose an additional symmetry, such as $\Z2$,
to stabilize it. Unlike the case of MFV, in BMFV, large flavor-violating effects are possible.
We have considered flavor, direct and indirect detection as well as relic
abundance constraints and on Figure~\ref{fig:summary_figure}
show the combined result for a representative BMFV model based on $\lambda_1$
couplings, which allows us to have $m_{\chi}, m_{\phi}$ 
in the several hundred GeV range. The most stringent constraints are those from the 
flavor-violating processes, and we can see that near term
new experiments will have the possibility to eliminate a
large region of parameter space below TeV for the BMFV model with $\lambda_1$, 
while keeping MFV-based scenario safe. 
We have also noted that if we are to assume that lepton-flavored DM
MFV based scenario can explain the 511 keV line splitting, following
the explanation of the 3.5 keV line in the recent literature, 
it is expected that there will be 
a sizable ($\sim$ keV) line-splitting induced near 511 keV.
In the BMFV case, due to the lack of constraints on the coupling matrix $\lambda$, 
the predicted splitting is not as definite. 
This splitting may be of interest for the future indirect-detection experiments.
Finally, the previous works
on MFV based lepton-flavored DM mainly considered hadron collider di-lepton
signals. On the other hand, our parton-level cross-section simulations as well as 
EFT considerations in the recent literature, show, 
that lepton $e^+ e^-$  collider constraints from mono-photon and fermion
pair production are stronger and are competitive among each other.
Finally, we have outlined possible implications for decaying DM,
if an additional stabilizing symmetry has not been imposed.

\appendix

\section{Calculating the relic abundance}
\label{app:calcrelic}

We briefly review relic abundance calculation (see \cite{Kolb:1990vq} for details)
and comment on co-annihilation effects. Consider several DM species $\chi_a, \chi_b$
with distinct masses $m_{\chi_a}, m_{\chi_b}$. For $\chi_a$,
 the relic abundance is determined by the self-annihilation process,  
$\chi_a \overline{\chi}_a \rightarrow l_i \overline{l}_j$, where $l_i$ is 
SM lepton of generation $i$.
This is done by solving the Boltzman equation for
evolution of the DM number density $n_a$
\begin{equation} \label{eq:boltzman}
\dfrac{dn_a}{dt} = - 3 H n_a - \langle \sigma v \rangle_{aa} [n_a^2 - (n^{eq})^2] ,
\end{equation}
where $H$ is the Hubble parameter governing expansion of the universe is
equal to $H = (8 \pi \rho/ 3 M_{Pl})^{1/2}$, $n^{eq}$ is the $\chi_a$ number
density at equilibrium,
$\langle \sigma v \rangle_{aa}$ is the thermally averaged
annihilation cross-section times the relative velocity. In the non-relativistic approximation,
$n^{eq}$ is given by
\begin{equation} \label{eq:boltzman_density}
n^{eq} \approx g \Big( \dfrac{m_{\chi_a} T}{ 2 \pi} \Big)^{3/2} e^{- m_{\chi} / T} ,
\end{equation}
where $T$ is the temperature and 
$g$ the number of degrees of freedom ($g = 4[2]$ for Dirac [Majorana] fermion).
Before the thermal averaging, in the non-relativistic limit 
one can expand the cross-section $\sigma = s + p v^2$,
thus separating $\langle \sigma v \rangle_{aa}$
into the non-relativistic ($s$-wave)  and relativistic  ($p$-wave) components.

For a single DM species present at the freeze-out,
the relic abundance in terms of $s$- and $p$-wave components
is approximately
\begin{equation}
\label{eq:relicabundance}
\Omega_{\chi} h^2 ~\approx~ \dfrac{1.07 \times 10^9}{ \text{GeV}}
\dfrac{1}{ M_{pl} \sqrt{g^{\ast}}} 
\Big(\dfrac{x_F}{s + 3 (p - s / 4) / x_F}\Big) ,
\end{equation}
where $\Omega_{\chi}$ is the present-day mass density divided by the
critical density $\rho_{c}$
\footnote{The critical mass density, which corresponds to a flat universe, is given by
$\rho_{c} = 3 H_0^2 M_{Pl}^2 / 8 \pi = 1.0539 \times 10^{-5} h^2\text{GeV cm}^{-3}$,
where $H_0 = 100 h~\text{km s}^{-1} \text{Mpc}^{-1}$ is the normalization
of the expansion rate.},
$h$ is the normalized expansion rate
and $x_F$ is the freeze-out temperature.
Here, $x_F$ is given by
\begin{equation}
\label{eq:xF}
x_F~=~\Bigg[ \dfrac{5}{8}  \sqrt{\dfrac{45}{8}} \Big(\dfrac{g}{\sqrt{g^{\ast}}}\Big)
\dfrac{M_{pl}}{\pi^3 } \dfrac{m_{\chi_a} (s + 6p / x_F)}{\sqrt{x_F}} \Bigg] ,
\end{equation}
where in the above $M_{pl}$ is the Planck scale given by
$M_{pl} = 1.22 \times 10^{19}$ GeV and 
$g^{\ast}$ is the temperature-dependent number of relativistic degrees 
of freedom at the freeze-out,
 taken here to be $g^{\ast} = 86.25$ (as in \cite{Bai:2013iqa}).
Fitting the relic abundance to the observed value 
of $\Omega_{\chi} h^2 ~=~ 0.1199 \pm 0.0027$ from
WMAP \cite{Hinshaw:2012aka} and Planck \cite{Ade:2013zuv}, 
we obtain the allowed dark matter parameter space.

For several DM species $\chi_{a}$ and $\chi_{b}$ with semi-degenerate masses, 
where $m_{\chi_{b}} \gtrsim m_{\chi_{a}}$, the relic density 
can be dominantly controlled by the co-annihilation process, 
$\chi_a \overline{\chi}_b \rightarrow l_{i} \overline{l}_{j}$ ($i, j =$ generation). 
Then, all $\chi_{b}$ decay into the stable DM candidate 
$\chi_{a}$, giving rise to the relic density in the universe today.

On the other hand, if the mass splitting between the species
is very small, $(m_{\chi_{a}} - m_{\chi_{b}}) < x_F / 20$, 
co-annihilation effects become important. Assuming small splitting and
that flavor changing processes $x_a l_i \rightarrow x_b l_j$
occur fast, all of the states are present in the freeze-out and co-annihilation dominates
\footnote{See \cite{Edsjo:1997bg, Ellis:1999mm} for co-annihilation 
effects within SUSY-based settings with neutralino DM.}.
The Boltzman equation has the same form as for single DM species
and the relic abundance will not be strongly affected.
At this point it is useful to define the effective 
cross-section, which for Dirac fermion DM as
considered within this work,
reads \cite{Srednicki:1988ce}
\begin{equation}
\langle \sigma v \rangle_{eff} = \dfrac{1}{2} \langle \sigma v \rangle .
\end{equation}
Then, the effects of co-annihilation are taken into account \cite{Griest:1990kh}
(see also \cite{Servant:2002aq}) as
\begin{equation}
\sigma_{eff}  = \sum_{i,j}^N \sigma_{i j} \Big( \dfrac{g_i g_j}{g_{eff}^2} \Big)
 \Big[ 1 + \Delta_i \Big]^{3/2} \Big[1 + \Delta_j \Big]^{3/2} e^{- x (\Delta_i + \Delta_j)}  ,
\end{equation}
where $\Delta_i = (m_i - m_1) / m_1$ is the mass difference between 
the $1^{st}$ and $i$'th DM particle,
$g_i$ are $i$'th relativistic d.o.f
and $g_{eff}$ are the effective d.o.f., given by
\begin{equation}
g_{eff} = \sum_{i = 1}^{N} g_i \Big[ 1 + \Delta_i \Big]^{3/2} e^{- x \Delta_i} .
\end{equation}
From the above, for highly degenerate DM with $m_i \approx m_j$,
as we have assumed for out model,
we can take $\Delta_i = 0$, resulting in
\begin{equation}
\sigma_{eff} = \sum_{i,j}^{N} \sigma_{i j} \Big(\dfrac{g_i g_j}{\sum_{a} g_a^2}\Big) .
\end{equation}

\acknowledgments

We would like to thank Shigeki Matsumoto, 
Satyanarayan Mukhopadhyay and Tim Tait for useful comments
regarding DM collider searches and Michael Ratz for useful discussions.  
V.T. would also like to thank Kavli IPMU, U. of Tokyo for hospitality, where part of the work was done.
M.-C.C. acknowledges the Aspen Center for Physics for hospitality and support. 
The work of M.-C.C. was supported, in part, by the U.S. National Science
Foundation under Grant No. PHY-0970173.  The work of V.T.\ was supported, in
part,  by the U.S.\ Department of Energy (DOE) under Grant No.\ DE--SC0009920. 
This work was partially performed at the Aspen Center for Physics, 
which is supported by National Science Foundation grant PHY-1066293. 

\bibliography{flavorDMbib}
\addcontentsline{toc}{section}{Bibliography}
\end{document}